\documentstyle[aps,preprint,tighten,epsf]{revtex}
\begin{document}
\draft
\title{
The Perturbed Static Path Approximation at Finite Temperature: 
Observables and Strength Functions}
\author{H. Attias and Y. Alhassid}
\address{Center for Theoretical Physics, Sloane Physics Laboratory, 
          Yale University, New Haven, Connecticut 06520}
\date{\today}
\maketitle

\begin{abstract}
\narrowtext
We present an approximation scheme for calculating observables and
strength functions of finite fermionic systems at  finite temperature
 such as hot nuclei.  The approach is formulated within the framework of 
the Hubbard-Stratonovich transformation and goes beyond the static path
 approximation and the RPA by taking into account small amplitude
 time-dependent
 fluctuations
around each static value of the auxiliary fields.  We show that this perturbed 
static path approach can be used systematically to 
obtain good approximations for observable expectation values and for
low moments of the strength function. The approximation for the strength 
function itself, extracted by an analytic continuation from the imaginary-time 
response function, is not always reliable, and we discuss the origin of the
discrepancies and possible improvements. Our results are 
tested in a solvable many-body model.
\end{abstract}
\narrowtext

\section{Introduction}

 Mean-field approximations\cite{ManyBody,ManyPart}, such as the finite
temperature Hartree-Fock (HF),  are standard for describing nuclei at finite
 temperature. Collective excitations of the system
are  obtained by linearizing the time-dependent HF equations around the
static HF solution, leading to the finite temperature random phase
 approximation (RPA) \cite{FTRPA}.  
 However, this treatment is 
inadequate for situations where various nuclear configurations compete and
 have comparable free energies, which causes observables to 
fluctuate widely about their mean values.  This is the case, for example, in
the vicinity of a shape transition from a spherical to a deformed nucleus,
where various discontinuities predicted in the mean-field approach  are  
smoothed in the finite nuclear system by the presence of fluctuations. 
Large fluctuations in the nuclear shape also play an important role in  
 giant dipole resonances (GDR) whose frequency is strongly coupled to the 
quadrupole deformation.  
In the adiabatic approximation, the observed GDR strength function is obtained 
by integrating  the strength function  that corresponds to each quadrupole 
configuration  over all possible configurations weighted by their respective
 Boltzmann factor \cite{StochGDR}.  

It has been shown \cite{ManyPart,AuxField} that a systematic description
of fluctuations can be obtained in the framework of the auxiliary-field 
path integral (Hubbard-Stratonovich transformation \cite{HubbStrat}).
 In this framework the
 original many-body propagator (in imaginary time) is decomposed 
into a superposition of one-body propagators describing
 non-interacting fermions moving in a fluctuating external time-dependent 
potential (auxiliary fields).
The expectation value of any observable is represented as a weighted average 
of its expectation values in the corresponding non-interacting systems.  
 Applying the method of
steepest descent one can obtain self-consistent mean-field approximations
 at finite  temperature. 
By taking into account small amplitude time-dependent fluctuations of the 
auxiliary fields around the  mean-field solution in the Gaussian
 approximation one rederives the RPA \cite{AuxField,AuxFieldArb}. 

 The path integral
representation also constitutes a starting point for exact numerical solution 
of the many-body problem where the auxiliary field integration is performed by 
 Monte Carlo techniques that were recently developed for strongly correlated 
electron systems \cite{LG} and  for the interacting nuclear shell model
 \cite{ShellMC}.  In the latter case a practical solution to the Monte Carlo 
sign problem \cite{Sign}, which is generic to all fermionic systems,
 enables  the study of nuclear properties in medium and heavy mass nuclei using
realistic  effective  interactions.
The auxiliary field formulation also suggests new approximation schemes
 which are
non-perturbative, the simplest of which is the static path approximation (SPA) 
\cite{SPAbcs} --\cite{SPAlip}. Here, the path 
integral is approximated by summing over the time-independent fields only,
 weighted 
by the appropriate Boltzmann factor.  This amounts to averaging over all 
possible static mean-field configurations rather than the self-consistent ones
alone. The SPA has been used to calculate free energies and level
 densities in 
nuclei and was found to be superior to the mean-field approximation.
 In particular, it accounts for the enhancement of level density due to thermal
 fluctuations of the shape \cite{SPAnuc}.
 More recently the approximation has  also been applied to the
calculation of strength functions \cite{SPAforG}. At high temperatures the 
SPA partition function approaches the exact result. 
However, as the temperature decreases the SPA
becomes inaccurate since time-dependent fluctuations about the mean-field 
configurations can no longer be neglected. This is  manifested especially in 
the strength function where even its first moment is significantly 
underestimated \cite{SPAforG}.

Recently, a method to improve the SPA 
 has been proposed for the partition function
\cite{PSPAnucbcs,PSPAforZnuc,PSPAforZ,PSPAforZbcs}. 
 In this approach contributions from
time-dependent fields in the neighborhood of each static field are
incorporated perturbatively to the second order in their amplitudes.
When one considers
such time-dependent fluctuations around the equilibrium configuration only,
one obtains the RPA corrections to the partition function. However,
in the proposed approach such small amplitude time-dependent fluctuations
are taken around each static configuration of the auxiliary fields and the
 static integration is still fully retained. 
At even lower temperatures some of these time-dependent fluctuations
 can become unstable and the approximation breaks down. However, this
happens only at temperatures below the phase transition, when the
 temperature drops below the largest imaginary RPA frequency. 
We remark that as a saddle point
 develops in the nuclear free energy surface below the transition
 temperature, it is still
possible that the imaginary RPA frequencies are small enough in magnitude
 that the time-dependent fluctuations are all stable (although there is an 
unstable direction in the static free energy surface). This
approximation scheme has been applied to free energy and level density 
calculations in simple models and shown to work well
 down to  low temperatures. 

It is interesting to investigate
whether incorporating small time-dependent fluctuations  provides a
significant improvement over the SPA evaluation of quantities 
other than the partition function. 
This paper will  
explore the  validity and applicability  of this scheme, which we 
term the perturbed static-path approximation (PSPA), 
for the calculation of expectation values of observables at finite temperature
 as well as 
of strength functions. Previous work employed a formulation of the PSPA
based on ordinary quantum-mechanical perturbation theory
 \cite{PSPAforZnuc,PSPAforZ},
which was specialized for free energy calculations and is not easily extended
to other quantities. We therefore reformulate it in a general way, using 
many-body 
methods, and apply it to the calculation of observables and strength 
functions. By testing the PSPA in a
solvable model we find that the PSPA results  agree closely  with the exact
solution for observable expectation values and for low moments
of the strength function. The strength function itself is approximated well
at high temperatures and also at low temperatures  for a certain regime of the
 model's parameters, although the low-temperature results generally do not
improve much on the SPA. We discuss possible improvements.

This paper is organized as follows. In Section II we review the auxiliary-field
path integral formalism and in Section III we formulate the PSPA for the 
partition
function. Section IV discusses the application of the PSPA to the calculation of
 expectation 
values of one- and two-body operators. Finally, in Section V we present a
treatment of the strength function and its lowest two moments in this
framework. We illustrate the PSPA and compare it  to other approximations
in  a simple model whose exact solution is given in Appendix B.

\section{Auxiliary Field Path Integral}
In this Section we briefly review the auxiliary-field path integral
representation of the imaginary-time evolution operator for an 
Hamiltonian with two-body interactions (Hubbard-Stratonovich transformation)
 \cite{ManyPart,AuxField}.
 We consider a system of interacting fermions, and assume for simplicity 
 an Hamiltonian of the form
\begin{eqnarray}\label{TwoBodyH}
   H=K-{1\over 2}\chi V^2 \;,
\end{eqnarray}
where $K$ and $V$ are one-body operators
\begin{eqnarray}\label{OneBodyKV}
   K=\sum\limits_{ij}k_{ij}a_i^\dagger a_j \;,\;\;\;\;\;
   V=\sum\limits_{ij}v_{ij}a_i^\dagger a_j \;.
\end{eqnarray}
  $a_i^\dagger$, $a_i$ are  creation and annihilation operators  for 
a set of single particle states $i$,
and $\chi$ is a coupling constant. Our results can be easily generalized
to a superposition of such separable interactions 
\begin{eqnarray}\label{ManyV}
   H=K-{1\over 2}\sum\limits_{\alpha=1}^q\chi_\alpha V_\alpha^2
   \;.
\end{eqnarray}
We remark  that an arbitrary two-body interaction 
\begin{eqnarray}\label{ArbitraryV}
   \sum\limits_{ijkl}u_{ijkl}a_i^\dagger a_j^\dagger a_l a_k
   &=&\sum\limits_{ij}\left(\sum\limits_k u_{ikjk}\right)a_i^\dagger a_j-
   \sum\limits_{ijkl}u_{iklj}a_i^\dagger a_j a_k^\dagger a_l
   \nonumber\\
   &\equiv&\sum\limits_{\alpha=1}^q{\tilde u}_\alpha\rho_\alpha-
   \sum\limits_{\alpha\beta=1}^q u_{\alpha\beta}\rho_\alpha\rho_\beta
   \;,\;\;\;\;\;
   \rho_\alpha=a_i^\dagger a_j
   \;,\;\;\;\;\;
   \rho_\beta=a_k^\dagger a_l
\end{eqnarray}
can be brought into the form (\ref{ManyV}) by diagonalizing $u_{\alpha\beta}$, 
with the number $q$ of separable interactions $V^2_\alpha$  equals at most 
to the square of the number of  single particle states.
In order to obtain a path integral representation of $U=\exp(-\beta H)$ we 
divide the imaginary-time interval $[0,\beta)$ into $N$ sub-intervals of length 
$\epsilon=\beta/N$ and write
\begin{eqnarray}\label{ImagTimeU}
   U=\prod\limits_{n=1}^N 
   e^{-K\epsilon+{1\over 2}\chi V^2\epsilon}
   =\left[\prod\limits_{n=1}^N e^{-K \epsilon}
   e^{{1\over 2}\chi V^2\epsilon}
   \times\left(1+{\cal O}(\epsilon^2)\right) \right]
   \;,
\end{eqnarray}
 At each time slice $n$ we use the identity 
\begin{eqnarray}\label{GaussIden}
   e^{\lambda\hat{a}^2}=
   \sqrt{\lambda\over\pi}\int\limits_{-\infty}^\infty d\xi 
   e^{-\lambda\xi^2}e^{\pm 2\lambda \xi\hat{a}}
   \;,
\end{eqnarray}
valid for any operator $\hat{a}$ with a bounded spectrum and 
$\lambda>0$ (we shall comment on the case $\lambda<0$ below), to replace
$e^{{1\over 2}\chi V^2\epsilon}$ by an integral over an auxiliary
variable $\xi_n$. This is the Hubbard-Stratonovich (HS) transformation 
\cite{HubbStrat}, resulting in
\begin{eqnarray}\label{UafterHS}
   U=\left({\chi\epsilon\over 2\pi}\right)^{N\over 2}
   \int\prod\limits_{n=1}^N d\xi_n
   \exp\left(-{1\over 2}\chi\epsilon\sum\limits_{n=1}^N \xi_n^2\right)
   \prod\limits_{n=1}^N e^{-K\epsilon+\chi\xi_n V\epsilon}
   \;.
\end{eqnarray}
In the limit 
$N\rightarrow\infty$, $\xi_n$ becomes a field $\xi(\tau=n\epsilon)=\xi_n$
and we obtain the continuous version of the HS transformation
\begin{eqnarray}\label{PathIntU}
   U=\int{\cal D}[\xi] \exp\left[-{1\over 2}\chi\int\limits_0^\beta 
   d\tau\xi^2\left(\tau\right)\right]U_\xi
   \;,
\end{eqnarray}
where $U_\xi$ is the (imaginary time) propagator for a time-dependent  one-body
Hamiltonian  $H_\xi=K-\chi\xi(\tau)V$
\begin{eqnarray}
U_\xi = {\rm T}\exp\left\{-\int\limits_0^\beta 
   d\tau\left[ K-\chi\xi\left(\tau\right)V \right]\right\}\;,
\end{eqnarray}
and T denote time ordering.
Eq.(\ref{PathIntU}) provides a representation of the many-body evolution 
operator  as an average over 
one-body evolutions $U_\xi$ which correspond to non-interacting particles
 moving
in a fluctuating time-dependent field $\xi(\tau)$, weighted by a Gaussian
 factor.
 Following \cite{FeynFourier} it is advantageous 
to describe the field $\xi(\tau)$ in terms of its  Fourier components,
\begin{eqnarray}\label{FourierComp}
   \xi(\tau=n\epsilon) =\sum\limits_{r=-(N-1)/2}^{(N-1)/2}
   \sigma_r e^{i\omega_r\tau}
   \;,
\end{eqnarray}
where $\sigma_{-r} =\sigma_r^\ast$ to keep $\xi(\tau )$ real and
$\omega_r =2\pi r/\beta$  are the 
Matsubara frequencies. Here we assume the number of time slices $N$ to be odd.
  Rewriting the functional integral over  the field
in terms of $\sigma_r$, we have 
\begin{eqnarray}\label{FourierPathU}
   U = \int{\cal D}[\sigma] 
   \exp\left(-{1\over 2}\chi\beta\sum\limits_r\mid\sigma_r\mid^2\right)
U_\sigma
   \;,
\end{eqnarray}
where
\begin{eqnarray}
U_\sigma = {\rm T}\exp\left[-\int\limits_0^\beta d\tau\left(K-\chi\sigma_0 V
   -\chi\sum\limits_{r\neq 0}\sigma_r e^{i\omega_r\tau}V\right)\right] \;.
\end{eqnarray}
The measure in (\ref{FourierPathU}) is given by
 \begin{eqnarray}\label{FourierMeas}
   {\cal D}[\sigma] =\left(\chi\beta\right)^{N\over 2}
   {d\sigma_0\over\sqrt{2\pi}}
   \prod\limits_{r>0}{d\sigma_r^\prime d\sigma_r^{\prime\prime}\over\pi}
   \;,
\end{eqnarray}
where $\sigma_r=\sigma_r^\prime+i\sigma_r^{\prime\prime}$. 
The one-body Hamiltonian $H_\sigma$ whose corresponding propagator is
 $U_\sigma$ separates into static and 
time-dependent parts
\begin{eqnarray}\label{StaticTdep}
   H_\sigma&=&h_0+h_1
   \;,\nonumber\\
   h_0&=&K-\chi\sigma_0 V
   \;,\;\;\;\;\;
   h_1=-\chi\sum\limits_{r\neq 0}\sigma_r e^{i\omega_r\tau}V
\;.
\end{eqnarray}
 It is useful to introduce the interaction picture representation
${\cal U}_\sigma$ of the one-body propagator $U_\sigma$ with respect 
to the static part of the Hamiltonian  $h_0$ 
\begin{eqnarray}\label{InterPic}
   U_\sigma&=&e^{-\beta h_0}
   {\rm T}\exp\left[\int\limits_0^\beta d\tau h_1\left(\tau\right)\right]
   \equiv e^{-\beta h_0}{\cal U}_\sigma
   \;,
\end {eqnarray}
where $h_1(\tau)=e^{\tau h_0}h_1 e^{-\tau h_0}$.
We then rewrite (\ref{FourierPathU}) as
\begin{eqnarray}\label{StatDynU}
   U=\sqrt{\chi\beta\over{2\pi}}
   \int d\sigma_0 e^{-{1\over 2}\chi\beta\sigma_0^2} e^{-\beta h_0} 
   \times\int{\cal D}^\prime [\sigma]
   \exp\left(-\chi\beta\sum\limits_{r>0}\mid\sigma_r\mid^2\right){\cal U}_\sigma
\end{eqnarray}
with the measure 
\begin{eqnarray}\label{DynMeas}
   {\cal D}^\prime[\sigma]=\left(\chi\beta\right)^{N-1\over 2}
   \prod\limits_{r>0}{d\sigma_r^\prime d\sigma_r^{\prime\prime}\over\pi} \;.
\end{eqnarray}
Eq. (\ref{StatDynU}) describes the evolution operator for the two-body
$H$ as a Gaussian-weighted average of one-body evolutions $e^{-\beta h_0}$
corresponding to a static field $\sigma_0$, multiplied by a correction factor 
which represents the contribution from  time-dependent fluctuations 
 of the $\xi$-field about 
$\sigma_0$. In fact, as we demonstrate below, any quantity of interest 
can be written 
as an integral over its static-field value times a correction factor.
 The objective of this paper
is to approximate this factor for various quantities by evaluating the
 small-amplitude fluctuation
contribution to the integral over $\sigma_r  (r>0)$ and to explore the validity
of this approximation.

For a superposition of separable interactions (\ref{ManyV}) one introduces
auxiliary-field 
variables $\sigma_r^\alpha$ corresponding to each $V_\alpha$ but the 
preceding
development remains unchanged. We point out that for $\chi<0$ in 
(\ref{TwoBodyH}) one should use
\begin{eqnarray}\label{GaussIdenNeg}
   e^{-\lambda\hat{a}^2}=
   \sqrt{\lambda\over\pi}\int\limits_{-\infty}^\infty d\xi 
   e^{-\lambda\xi^2}e^{\pm 2i\lambda \xi\hat{a}}
\end{eqnarray}
($\lambda>0$) instead of (\ref{GaussIden}). 

\section{Partition Function}
We now consider the partition function $Z \equiv {\rm Tr}e^{-\beta H}$. We 
work in the grand canonical ensemble and set the chemical potential 
$\mu=0$ to keep the notation
simple. It is convenient to choose a $\sigma_0$-dependent basis for the Fock 
space in which $h_0$ is diagonal:
\begin{eqnarray}\label{StatBasis}
   h_0=\sum\limits_i\epsilon_i(\sigma_0) a_i(\sigma_0)^\dagger a_i(\sigma_0)
   \;,\;\;\;\;\;
   V=\sum\limits_{ij}v_{ij}(\sigma_0) a_i(\sigma_0)^\dagger a_j(\sigma_0)
   \;.
\end{eqnarray}
We can write $Z$ using (\ref{FourierPathU}) as
\begin{eqnarray}\label{FourierPathZ}
   Z={\rm Tr}U=\int{\cal D}[\sigma] 
   \exp\left(-{1\over 2}\chi\beta\sum\limits_r\mid\sigma_r\mid^2\right) 
   {\rm Tr}U_\sigma
   \equiv\int{\cal D}[\sigma] e^{-\beta F(\beta;\sigma)}
   \;,
\end{eqnarray}
or in a form more convenient for our purpose (using (\ref{StatDynU})):
\begin{eqnarray}\label{StatDynZ}
   Z=\sqrt{\chi\beta\over{2\pi}}
   \int d\sigma_0 e^{-{1\over 2}\chi\beta\sigma_0^2}\zeta_0\zeta_0^\prime
   \equiv\sqrt{\chi\beta\over{2\pi}}\int d\sigma_0 e^{-\beta F_0(\beta;
\sigma_0)}
   \;.
\end{eqnarray}
Here
\begin{eqnarray}\label{StatZ}
   \zeta_0={\rm Tr}e^{-\beta h_0}=\prod\limits_i 
   \left[1+e^{-\beta\epsilon_i\left(\sigma_0\right)}\right]
\end{eqnarray}
is the partition function corresponding to the static part $h_0$ and 
\begin{eqnarray}\label{DynZ}
   \zeta_0^\prime=\int{\cal D}^\prime[\sigma]
   \exp\left(-\chi\beta\sum\limits_{r>0}\mid\sigma_r\mid^2\right)
   {1\over\zeta_0}{\rm Tr} \left(e^{-\beta h_0}{\cal U}_\sigma\right)
   =\int{\cal D}^\prime[\sigma]
   \exp\left(-\chi\beta\sum\limits_{r>0}\mid\sigma_r\mid^2\right)
   \langle{\cal U}_\sigma \rangle_0
\end{eqnarray}
is the correction factor to $\zeta_0$ due to the time-dependent fluctuations 
of the field about $\sigma_0$. We use the notation 
$\langle O\rangle_0 \equiv {\rm Tr}(e^{-\beta h_0}O)/\zeta_0$ to denote 
the thermal average of an observable $O$ with respect to $h_0$.
The effective static-field free energy $F_0$ is defined by (\ref{StatDynZ}) 
to be 
\begin{eqnarray}\label{Fzero}
   F_0(\beta;\sigma_0)={1\over 2}\chi\sigma_0^2
   -{1\over\beta}\log\zeta_0-{1\over\beta}\log\zeta_0^\prime
   \;.
\end{eqnarray}

The representation (\ref{StatDynZ}) of $Z$ is a starting point for various
 approximations. The mean-field approximation (MFA) 
is obtained when the contribution of the time-dependent paths to $Z$ is
neglected by setting $h_1=0$, implying $\zeta_0^\prime=1$, 
and the integration
over the static fields $\sigma_0$ is performed in the method of steepest 
descent. This amounts to 
approximating the path integral by the contributions of the static paths 
$\bar{\sigma}_0$ that minimize the free energy $F_0$ in (\ref{Fzero}). It is 
easy to show \cite{ManyPart,AuxField} that this minimization condition is 
\begin{eqnarray}\label{MinFzero}
   \bar{\sigma}_0 = \langle V \rangle_0 =
\sum\limits_{i}v_{ii}(\bar{\sigma}_0)f_i(\bar{\sigma_0})
   \;,
\end{eqnarray}
where $f_i=(1+e^{\beta\epsilon_i})^{-1}$ are the Fermi occupation numbers,
and that the solution $\bar{\sigma}_0$ of (\ref{MinFzero}) is the Hartree mean 
field. One can improve on the MFA result by performing the integration over
$\sigma_r$  in the expression  (\ref{DynZ}) for $\zeta_0^\prime$ also by
 steepest descent.  The saddle point is now given by
 $\bar{\sigma}_0, \sigma_r=0$
and one obtains the 
finite temperature RPA corrections to the partition function \cite{AuxField}. 
 
The static-path approximation (SPA) \cite{SPAbcs} -- \cite{SPAlip}
is obtained by again setting $h_1=0$ but now the integration over 
$\sigma_0$ in (\ref{StatDynZ}) is performed exactly, thus approximating
 the path integral by the contributions from all static paths:
\begin{eqnarray}\label{Zspa}
   Z^{(SPA)}=\sqrt{\chi\beta\over{2\pi}}
   \int d\sigma_0 e^{-{1\over 2}\chi\beta\sigma_0^2}\zeta_0
   =\sqrt{\chi\beta\over{2\pi}}
   \int d\sigma_0 e^{-{1\over 2}\chi\beta\sigma_0^2}
   \prod\limits_i 
   \left[1+e^{-\beta\epsilon_i\left(\sigma_0\right)}\right]
   \;.
\end{eqnarray}
The SPA is expected to become exact at high temperatures since
one can use the one time slice approximation in (\ref{UafterHS}) with  
an error of ${\cal O}(\beta^2)$ that vanishes as $T\rightarrow\infty$. 
This method is advantageous to the 
MFA since it takes into account exactly large amplitude static fluctuations
 around the mean field.
However, it neglects the time-dependent fluctuations which constitute
the RPA corrections. This shortcoming of the SPA can be remedied if the exact
integration over the static paths is supplemented by 
evaluation of $\zeta_0^\prime$ in the limit of small amplitude time-dependent
 fluctuations about each static value $\sigma_0$.
 This scheme, the perturbed
static-path approximation (PSPA), is the focus of our work. It was introduced 
in Refs.  \cite{PSPAnucbcs} - \cite{PSPAforZbcs} for the 
 partition function $Z$. Our approach is different and has the advantage that 
it can be generalized to the calculation of observables and response functions
as shown in the following sections.
We first illustrate our method by rederiving the corresponding expression for
 the partition function.  We  expand 
$ \log\langle{\cal U}_\sigma\rangle_0$ in (\ref{DynZ})
\begin{eqnarray}\label{ExpandLnU} 
   \log\langle{\cal U}_\sigma\rangle_0&=&
   \chi\sum\limits_{r\neq 0}\sigma_r 
   \int\limits_0^\beta d\tau e^{i\omega_r\tau}\langle V(\tau)\rangle_c
   \nonumber\\
   &+&{1\over 2}\chi^2\sum\limits_{rs\neq 0}\sigma_r\sigma_s
   \int\limits_0^\beta d\tau d\tau^\prime e^{i\omega_r\tau}
   e^{i\omega_s\tau^\prime} 
   \left[\langle{\rm T}V(\tau)V(\tau^\prime)\rangle_c
   -\langle V(\tau)\rangle_c\langle V(\tau^\prime)\rangle_c\right]
   \nonumber\\
   &+&{\cal O}\left(\sigma^3\right)
   \;,
\end{eqnarray}
with $V(\tau)=e^{\tau h_0}Ve^{-\tau h_0}$ in the interaction picture. The 
time-ordered averages are calculated using the finite-temperature Wick's 
theorem \cite{ManyBody,ManyPart} where the subscript $c$ means 
that only the connected diagrams should be summed up in the diagrammatic
 representation of  this expansion. 

The first term in (\ref{ExpandLnU}) vanishes since $\langle V(\tau)\rangle_c$ 
is $\tau$-independent. For the second term Wick's theorem gives
\begin{eqnarray}\label{VtauVtaup}
   \langle{\rm T}V(\tau)V(\tau^\prime)\rangle_c=
   -\sum\limits_{ij}v_{ij}v_{ji}g_i^0(\tau^\prime-\tau)g_j^0(\tau-\tau^\prime)
   \;,
\end{eqnarray}
where the unperturbed temperature Green's function $g_i^0$ is given by 
\begin{eqnarray}\label{TempGreenFn}
   g_i^0(\tau-\tau^\prime )= 
   -\langle a_i(\tau)a_i^\dagger(\tau^\prime)\rangle_0
   ={1\over\beta}\sum\limits_{k=-\infty}^\infty 
   {e^{-i\nu_k(\tau-\tau^\prime)}\over i\nu_k-\epsilon_i}
\end{eqnarray}
with frequencies $\nu_k=(2\pi +1)k/\beta$.
Using (\ref{VtauVtaup}-\ref{TempGreenFn}) we obtain for the double integral in 
(\ref{ExpandLnU})
\begin{eqnarray}\label{DoubleIntToSum}
   \int\limits_0^\beta d\tau d\tau^\prime\cdots=
   -\delta_{r,-s}\sum\limits_{ij}v_{ij}v_{ji}
   \sum\limits_{k=-\infty}^\infty{1\over i\nu_k-\epsilon_i} \;
   {1\over i\nu_k-(\epsilon_j-i\omega_r)}
   \;.
\end{eqnarray}
The infinite sum over $k$ in (\ref{DoubleIntToSum}) is calculated using the 
frequency summation technique \cite{ManyBody,ManyPart}.
The essence of this method lies in the observation that the points 
$z=i\nu_k$ are the poles of the function $-\beta/(e^{\beta z}+1)$ with residue 
of one and the sum therefore equals the contour integral 
\begin{eqnarray}\label{SumToContourInt}
   \sum\limits_{k=-\infty}^\infty{1\over i\nu_k-\epsilon_i} \;
   {1\over i\nu_k-(\epsilon_j-i\omega_r)}=
   {1\over 2\pi i}\oint\limits_{\cal C} dz \; {-\beta\over e^{\beta z}+1} \;
   {1\over z-\epsilon_i} \; {1\over z-(\epsilon_j-i\omega_r)}
   \;,
\end{eqnarray}
where ${\cal C}$ encircles the imaginary axis. This contour can be continuously 
transformed into a circle centered at the origin of an arbitrarily large 
radius, which is deformed at two places to include the poles at $z=\epsilon_i$ 
and $z=\epsilon_j-i\omega_r$. The residue theorem then gives  
\begin{eqnarray}\label{DoubleIntFinal}
   \int\limits_0^\beta d\tau d\tau^\prime\cdots=
   -\beta\delta_{r,-s}\sum\limits_{ij}v_{ij}v_{ji}
   {f_i-f_j\over\Delta_{ij}+i\omega_r}
   \;,
\end{eqnarray}
with $\Delta_{ij}=\epsilon_i-\epsilon_j$. Using (\ref{DoubleIntFinal}) in 
(\ref{ExpandLnU}) we have
\begin{eqnarray}\label{AverageU}
   \langle{\cal U}_\sigma\rangle_0=
   \exp\left(-\chi\beta\sum\limits_{r>0}a_r\mid\sigma_r\mid^2\right)
   \;,
\end{eqnarray}
with
\begin{eqnarray}\label{AsubR}
   a_r(\sigma_0)= \chi \sum\limits_{ij}v_{ij}v_{ji}
   {f_i-f_j\over\Delta_{ij}+i\omega_r}
 = \chi\sum\limits_{ij}v_{ij}v_{ji}
   {\left(f_i-f_j\right)\Delta_{ij}\over\Delta_{ij}^2+\omega_r^2}
   \;.
\end{eqnarray}
The correction factor $\zeta_0^\prime$ in (\ref{DynZ}) is now given by a 
Gaussian integral and we obtain
\begin{eqnarray}\label{DynZGauss}
   \zeta_0^\prime=\prod\limits_{r>0}\left(1+a_r\right)^{-1}
   =\prod\limits_{r>0}
   {\prod\limits_{ij}{^\prime}\left(\omega_r^2+\Delta_{ij}^2\right)\over
    \prod\limits_\nu\left(\omega_r^2+\Omega_\nu^2\right)}
   ={\prod\limits_{ij}{^\prime}{1\over\Delta_{ij}}
                             \sinh{\beta\Delta_{ij}\over 2}\over
     \prod\limits_\nu{1\over\Omega_\nu}\sinh{\beta\Omega_\nu\over 2}}
   \;.
\end{eqnarray} 
The second equality in (\ref{DynZGauss}) defines the frequencies 
$\Omega_\nu(\beta,\sigma_0)$ through \cite{PSPAforZ}
\begin{eqnarray}\label{DefOmegaNu}
   1+a_r=
   {\prod\limits_{ij}{^\prime}(\omega_r^2+\Delta_{ij}^2)
   +2\chi\sum\limits_{ij}{^\prime}v_{ij}v_{ji}(f_i-f_j)\Delta_{ij}
   \prod\limits_{kl\neq ij}{^\prime}(\omega_r^2+\Delta_{kl}^2)
   \over
   \prod\limits_{ij}{^\prime}(\omega_r^2+\Delta_{ij}^2)}
   \equiv{\prod\limits_\nu(\omega_r^2+\Omega_\nu^2)
   \over
   \prod\limits_{ij}{^\prime}(\omega_r^2+\Delta_{ij}^2)} \;,
\end{eqnarray}
where the prime in $\prod\limits_{ij}{^\prime}$ and $\sum\limits_{ij}{^\prime}$
restricts the product or sum to pairs $(i,j)$ that satisfy $i<j$ and 
$\Delta_{ij}\neq 0$. Note that there are as many $\Omega_\nu$ in 
 the numerator of  (\ref{DefOmegaNu}) as there are $\Delta_{ij}$ 
in the denominator. The third equality in 
(\ref{DynZGauss}) uses the infinite product representation 
$\sinh x=x\prod\limits_{r>0}(1+ x^2 / \pi^2 r^2)$.  
Together with (\ref{StatDynZ}) and (\ref{StatZ}) we finally have
\begin{eqnarray}\label{Zpspa}
   Z^{(PSPA)}=\sqrt{\chi\beta\over{2\pi}}
   \int d\sigma_0 e^{-{1\over 2}\chi\beta\sigma_0^2}
   \prod\limits_i 
   \left(1+e^{-\beta\epsilon_i\left(\sigma_0\right)}\right) 
   {\prod\limits_{ij}{^\prime}{1\over\Delta_{ij}}
                            \sinh{\beta\Delta_{ij}\over 2}\over
    \prod\limits_\nu{1\over\Omega_\nu}\sinh{\beta\Omega_\nu\over 2}}
   \;.
\end{eqnarray}
 This is a closed-form expression which corresponds to the 
limit $N \rightarrow \infty$ ($N$ is the number of imaginary-time slices, see 
(\ref{ImagTimeU})) since the infinite product in (\ref{DynZGauss}) has been 
performed exactly. Hence the PSPA result does not contain errors originating
from a discretization of $[0,\beta)$ into sub-intervals of a finite length, 
as is the case in a Monte Carlo evaluation of the path integral.

 $\omega=\pm \Omega_\nu$ are the roots of $1 + \sum\limits_{ij}v_{ij}v_{ji}
   (f_i-f_j) / (\Delta_{ij}+i\omega) =0$.
It can be shown \cite{PSPAforZ}
that for the value $\sigma_0=\bar{\sigma_0}(\beta)$ which minimizes
$F_0(\beta;\sigma_0)$ (see Eqs. (\ref{Fzero})  and (\ref{MinFzero})),
 these roots $\pm\Omega_\nu$ are the 
RPA frequencies, i.e. the frequencies of small amplitude  oscillations
 around the  equilibrium configuration. For an arbitrary static $\sigma_0$, 
the frequencies $ \omega = \pm \Omega_\nu$ solve  the generalized
 finite temperature RPA equations obtained by replacing  the equilibrium
configuration by the arbitrary  $\sigma_0$ (see  Appendix A).
 We note that the Gaussian approximation in (\ref{DynZ}) leads to a convergent 
integral only
if $1+a_r > 0$ for all $r$, i.e. $-\Omega_\nu^2 < \omega^2_r$ for all $r$ and 
$\nu$.
 When all RPA frequencies are real, these conditions are always met. 
In particular
this is the case if $\sigma_0$ is a local minimum of the static free energy 
 i.e. a stable mean-field configuration $\bar{\sigma}_0$
 (Thouless theorem \cite{Thou}). 
At zero temperature $\omega_r=0$ and an imaginary RPA frequency
would lead to
a breakdown of the Gaussian approximation. 
However, at finite temperature imaginary RPA frequencies do not necessarily
lead to instability.  If the largest modulus of all imaginary RPA frequencies is
below $\omega_1=2\pi T$, then the quadratic fluctuations in $\sigma_r$
 are still stable.
An instability in the $\sigma_1$ direction occurs when the magnitude of
 one of the imaginary
 RPA frequencies crosses $2\pi T$. 
The RPA frequencies depend both on $\beta$ and $\sigma_0$.
 In particular, for any temperature $\beta^{-1}$ there exists  a static field
$\sigma_0=\sigma_0^\prime(\beta)$
such that 
$\Omega_\nu^2(\beta,\sigma_0^\prime)=-\omega_1^2=-(2\pi/\beta)^2$ for some 
$\nu$, causing the correction factor $\zeta_0^\prime$ to diverge.
 Note that at $T=0$ instability occurs as 
 soon as some $\Omega_\nu$ becomes imaginary whereas at
$T>0$ an instability develops only for a sufficiently imaginary $\Omega_\nu$.
These instabilities can be ignored when they occur 
at fields for which the static free energy is large.  In practical applications
 the PSPA breaks down only at
temperatures that are significantly below the shape transition temperature, when
the saddle point in the static free energy surface becomes unstable to small
 amplitude time-dependent fluctuations. 

In the case (\ref{ManyV}) where the interaction is a sum of several  separable
 interactions we have
an auxiliary-field variable $\sigma_r^\alpha$ for each $V_\alpha$. Assuming
$\chi_\alpha=\chi$ (this is always possible by redefining $V_\alpha$),   
expression (\ref{AverageU}) becomes
\begin{eqnarray}\label{AverageUManyV}
   \langle{\cal U}_\sigma\rangle_0=
   \exp\left(-\chi\beta \sum\limits_r \sum\limits_{\alpha\gamma}
   {\sigma_r^\alpha}^\ast a_r^{\alpha\gamma} \sigma_r^\gamma \right)
\end{eqnarray}
with
\begin{eqnarray}\label{ABsubR}
   a_r^{\alpha\gamma}(\sigma_0)=\chi\sum\limits_{ij}v_{ij}^\alpha v_{ji}^\gamma
   {\left(f_i-f_j\right)\over\Delta_{ij}+ i\omega_r}
   \;.
  \end{eqnarray}
The correction factor $\zeta_0^\prime$ is then generalized from
(\ref{DynZGauss}) into
\begin{eqnarray}\label{DynZGaussManyV}
   \zeta_0^\prime=\prod\limits_{r>0}\det\left(1+a_r\right)^{-1} 
\end{eqnarray}
where $a_r$ is the $q$-dimensional matrix  defined in (\ref{ABsubR}).
 As shown in Appendix A
\begin{eqnarray}\label{DetManyV}
\det(1+a_r) = {\prod\limits_\nu(\omega_r^2+\Omega_\nu^2)
   \over
   \prod\limits_{ij}{^\prime}(\omega_r^2+\Delta_{ij}^2)} \;,
\end{eqnarray}
where $\Omega_\nu$ are again the RPA frequencies at finite temperature.
We then obtain for the PSPA partition function an expression similar to
 (\ref{Zpspa}),
 except that the integral over $\sigma_0$ is replaced by an integration over $q$
static fields $\sigma^\alpha_0$.

   When one or more of the $\chi_\alpha$ in (\ref{ManyV}) are negative and a
 representation of the type (\ref{GaussIdenNeg}) is used in the HS
 transformation, the one-body Hamiltonian
$h_0$ in (\ref{StatBasis})  becomes non-hermitean, so that its eigenvalues 
$\epsilon_i(\sigma_0)$
  are in general complex and the associated one-body  partition function
 can be negative. It was recently
 pointed out \cite{RC97} that  the static fields which correspond to these
 ``repulsive'' terms in the interaction do not represent large amplitude
 thermal fluctuations, and can simply be treated in a saddle point 
approximation.  This amounts to keeping the exact integration over the static
 ``attractive'' fields but taking the mean-field solution for the
static ``repulsive'' fields for every configuration of the
``attractive'' fields  in the integrand.

   An alternative way  to approach ``repulsive''  interactions  is
to subtract from the Hamiltonian a term proportional to $\hat{N}^2$ ($\hat{N}$
 is the particle number operator) with a coefficient large enough so that the
 two-body part of the Hamiltonian becomes a negative-definite
 quadratic form (as in the case where all $\chi_\alpha$ in (\ref{ManyV})
 are positive).   However, by doing this we also  increase
 the magnitude of   ``attractive'' terms and 
 it remains to be investigated whether the SPA and PSPA would still work well 
for such modified Hamiltonians.  Note that the subtracted term is just a 
constant for a fixed number of particles.
 
To illustrate our results we apply the formalism to a simple many-body model 
(a variant of the Lipkin model \cite{Lipkin}) based on a $U(2)$ algebra which is
 defined and solved in Appendix B. 
At each  $\sigma_0$ we have
\begin{eqnarray}\label{StaticTdepLip}
   h_0=2\epsilon J_z-2\chi\sigma_0 J_x
   \;,\;\;\;\;\;
   h_1=-2\chi\sum\limits_{r\neq 0}\sigma_r e^{i\omega_r\tau}J_x
   \;.
\end{eqnarray}
Thus the single-particle Hamiltonian corresponding to $h_0$ has two $g$-fold
degenerate levels $\epsilon_i=\pm\bar{\epsilon}$ where 
$\bar{\epsilon}=\sqrt{\epsilon^2+\chi^2\sigma_0^2}$, so 
$\Delta_{ij}=0,\pm 2 \bar{\epsilon}$. The matrix 
corresponding to $V=2J_x$ is block-diagonal with $g$ $\; 2\times 2$-blocks
\begin{eqnarray}\label{VmatLip}
   v_{ij}=\left(\begin{array}{c}
   -\sin 2\phi \;\;\;\;\; \cos 2\phi \\
   \;\; \cos 2\phi \;\;\;\;\; \sin 2\phi
   \end{array}\right)
   \;,\;\;\;\;\;
   \sin 2\phi={\chi\sigma_0\over\bar{\epsilon}}
   \;,\;\;\;\;\;
   \cos 2\phi={\epsilon\over\bar{\epsilon}}
\end{eqnarray}
and the matrix 
\begin{eqnarray}\label{Pdiag}
   p_{ij}=\left(\begin{array}{c}
   \;\; \cos\phi \;\;\;\;\; \sin\phi \\
   -\sin\phi \;\;\;\;\; \cos\phi
   \end{array}\right)
\end{eqnarray}
diagonalizes the $2\times 2$-blocks of $h_0$. For $\zeta_0$ and 
$\zeta_0^\prime$ we obtain
\begin{eqnarray}\label{LipkinZ}
   \zeta_0=\left(2\cosh{\beta\bar{\epsilon}\over 2}\right)^{2g}
   \;,\;\;\;\;\;
   \zeta_0^\prime={\Omega\over 2\bar{\epsilon}} \;
   {\sinh\beta\bar{\epsilon}\over\sinh{\beta\Omega\over 2}}
   \;,
\end{eqnarray}
where $\Omega^2=4\bar{\epsilon}^2-(4g\epsilon^2\chi/\bar{\epsilon})
                                  \tanh(\beta\bar{\epsilon}/2)$,
so there are only two RPA frequencies $\pm\Omega$. 
The saddle-point  mean-field equation (\ref{MinFzero}) is
\begin{eqnarray}\label{MinFzeroLip}
   \tanh{\beta\bar{\epsilon}\over 2}={\bar{\epsilon}\over g\chi}
\end{eqnarray}
and has a single spherical stable solution $\sigma_0=0$ for $\beta<\beta_c$,
 which becomes unstable  for $\beta>\beta_c$ and bifurcates into two 
 deformed minima  $\sigma_0=\pm\bar{\sigma}_0(\beta)$. Here $\beta_c$ is the
 transition temperature
\begin{eqnarray}\label{BetaCLip}
   \beta_c={1\over\epsilon}\log{\kappa+1\over\kappa-1}
   \;,\;\;\;\;\;
   \kappa={g\chi\over\epsilon}
   \;.
\end{eqnarray}
The dimensionless parameter $\kappa$ thus determines the mean-field
 behavior of the model. For $\kappa<1$ there is no phase transition and 
$\sigma_0=0$ is the only solution at all temperatures.

\begin{figure}[p]
\vspace{1 cm}

\centerline{\epsffile{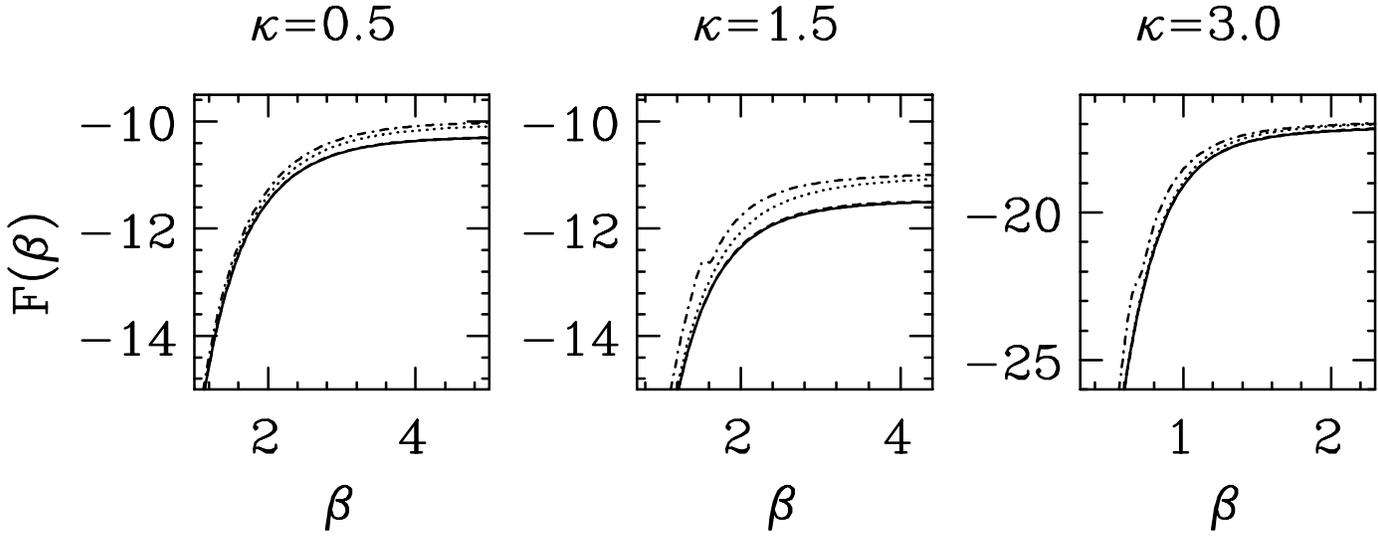}}

\vspace*{-8 cm}

\caption{ 
Free energy $F(\beta)=-\beta^{-1} \log Z$ as a function of $\beta$ for 
different values of $\kappa$ (see  (\protect\ref{BetaCLip})). The SPA
 (dotted) and PSPA (dashed)
results are obtained using (\protect\ref{LipkinZ}) in Eqs. 
(\protect\ref{Zspa}) and
(\protect\ref{Zpspa}). The MFA result (dashed-dotted) is given by a 
steepest-descent treatment of the SPA integral. The exact result (solid) is
calculated from (\protect\ref{a5}).
} 
\label{Free}
\end{figure}

In Fig. \ref{Free} we present the MFA, SPA and PSPA results for the free energy
$F(\beta) \equiv -\beta^{-1}\log Z$ in our $U(2)$ model using 
$\epsilon=1$ and $g=10$
and compare them against the exact result. We consider three cases 
characterized by different values of $\kappa$. For $\kappa=0.5$ there is no 
mean-field transition whereas for $\kappa=1.5$ and $\kappa=3.0$ the signature 
of a transition at temperatures $\beta_c=1.61$ and $\beta_c=0.693$, 
respectively, is seen in the mean-field curve. 
The SPA overestimates the 
exact result especially at low $T$ whereas the PSPA is quite accurate
at all temperatures. All three approximations converge to the exact result as
$T\rightarrow\infty$. We find that for $\beta$ 
sufficiently small the divergence of $\zeta_0^\prime$ occurs only at a very 
large $\sigma_0=\sigma_0^\prime(\beta)$, where the other factors in the 
integrand
of (\ref{Zpspa}) are vanishingly small, and therefore does not constitute a 
practical problem. The breakdown of the PSPA occurs when $\beta$ becomes
large enough such that $\sigma_0^\prime(\beta)\sim 0$. This happens only at
temperatures far below the transition temperature $T_c=1/\beta_c$.

\begin{figure}[p]

\vspace*{1 cm}

\centerline{\epsffile{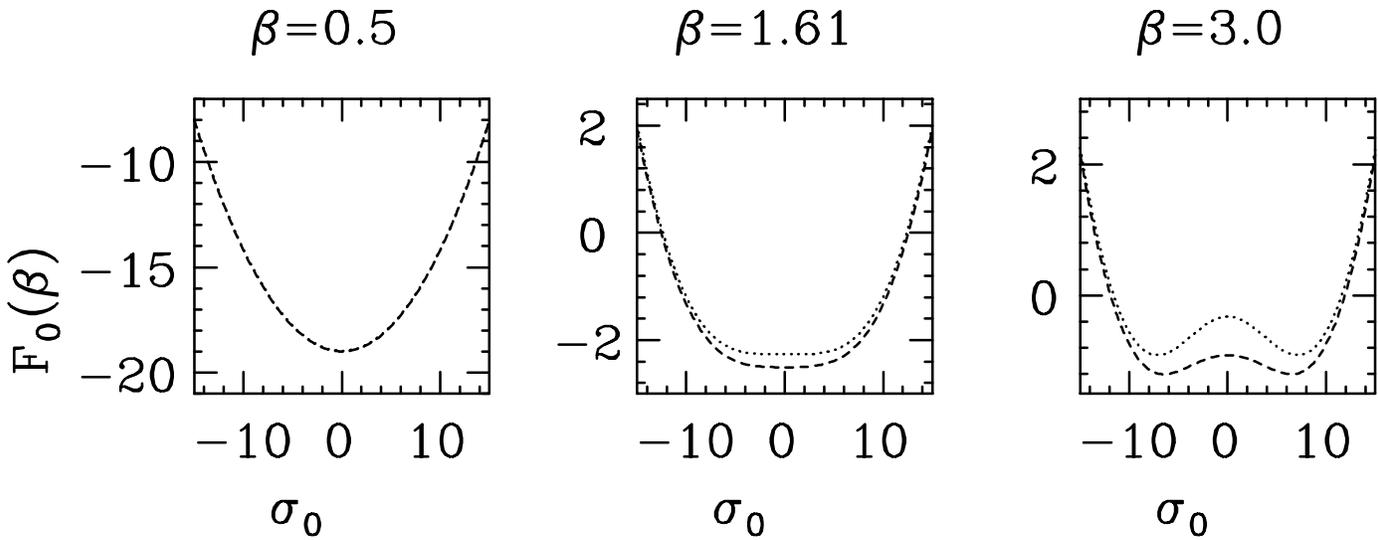}}

\vspace*{-8 cm}

\caption{
Effective static-field free energy $F_0(\beta; \sigma_0)$ 
(\protect\ref{Fzero}) as a 
function of $\sigma_0$ in the SPA (dotted) and PSPA (dashed) at different 
temperatures $\beta$. Shown is the case $\kappa=1.5$ where the 
mean-field phase-transition occurs at $\beta_c=1.61$.
} 
\label{Free0}
\end{figure}

In order to study the effect of  the time-dependent fluctuations we compare in 
Fig. \ref{Free0}  the effective  free energy 
$F_0(\beta;\sigma_0)$ in (\ref{Fzero}) as a function of 
the static field $\sigma_0$  with the SPA free energy ($\zeta^\prime_0=1$
in  (\ref{Fzero})) at different temperatures. We consider the case 
$\kappa =1.5$. 
Above the mean-field transition temperature $\beta_c=1.61$, 
$\zeta_0^\prime\approx 1$ and
the two approximations yield  similar results. As the temperature is
lowered, time-dependent fluctuations deepen the free-energy minimum
 at $\sigma_0=0$.  
 Below $T_c$ these fluctuations also lower the barrier between the two 
mean-field configurations $\sigma_0=\pm\bar{\sigma}_0$. Thus the PSPA result for
 the free energy
 improves significantly the SPA result especially below the transition 
temperature
(see the middle panel of Fig. \ref{Free}).  The MFA result  for the free energy 
$F(\beta)$ is the most  inaccurate
 especially near the transition where the free energy has
comparable values over a broad range of  configurations $\sigma_0$.

  We remark that in spite of its simplicity, the $U(2)$ model has features 
that are 
generic to more realistic nuclear interactions (e.g. quadrupole interaction).
 In particular, the phase transition in its mean-field theory is analogous to
 the shape transitions from deformed to spherical shapes
 that occur in deformed nuclei \cite{Al91}.   We therefore expect the
 PSPA to improve significantly on the SPA also for more realistic
 nuclear shell model interactions. 
 For  interactions that include both pairing and multipole 
components (e.g. pairing plus quadrupole model),  the SPA 
  works better  when a mixed pairing-density
 decomposition is used  rather than just a pure density 
decomposition\cite{ShellMC}. A mixed decomposition is thus preferable 
 when the PSPA is applied to such interactions.

\section{Thermal Expectation Values of Observables} 

In this Section we consider an observable $O$ and treat its expectation value
at finite temperature $\langle O\rangle=
{\rm Tr}(e^{-\beta H}O)/{\rm Tr}(e^{-\beta H})$ in the
framework of the PSPA. Although we discuss observables of the form
$O=D$ and $O=D^2$ for a one-body operator D, any $n$-body $O$ can be 
treated in a similar fashion.

\subsection{One-Body Observables}
In constructing an auxiliary-field path integral representation for 
$\langle D\rangle$ we can use Eqs. (\ref{FourierPathU}) and 
(\ref{FourierPathZ}) to get
\begin{eqnarray}\label{FourierPathD}
   \langle D\rangle={\int{\cal D}[\sigma] e^{-\beta F(\beta;\sigma)}
 \langle D \rangle_\sigma\over
                     \int{\cal D}\sigma e^{-\beta F_\sigma}}
   \;,\;\;\;\;\;
   \langle D \rangle_\sigma={{\rm Tr}(U_\sigma D)\over{\rm Tr}U_\sigma}
   \;.
\end{eqnarray}
Alternatively, to facilitate our calculations we can use 
(\ref{StatDynU}) and (\ref{StatDynZ}) to write
\begin{eqnarray}\label{StatDynD}
   \langle D\rangle={\int d\sigma_0 e^{-\beta F_0}D_0\over
                     \int d\sigma_0 e^{-\beta F_0}}
\end{eqnarray}                    
where
\begin{eqnarray}\label{Dzero}
   D_0={1\over\zeta_0^\prime}\int{\cal D}^\prime [\sigma]  
   \exp\left(-\chi\beta\sum\limits_{r>0}\mid\sigma_r\mid^2\right)
   \langle{\cal U}_\sigma D\rangle_0
   \;.
\end{eqnarray}
The effective static-field free energy $F_0(\beta;\sigma_0)$, defined in
 (\ref{Fzero}),
has been calculated in the previous Section 
(see Eqs.  (\ref{StatZ}) and (\ref{DynZGauss})).
Our aim here is to calculate $D_0$. In a diagrammatic representation of the
perturbation series for $\langle{\cal U}_\sigma D\rangle_0$ the
 connected diagrams
can be easily shown to factor out at each order \cite{ManyBody,ManyPart}:
\begin{eqnarray}\label{FactorDconn}
   \langle{\cal U}_\sigma D\rangle_0=
   \langle{\cal U}_\sigma\rangle_0\langle{\cal U}\sigma D\rangle_c
   \;.
\end{eqnarray}
Thus we can write $D_0$ itself as a weighted average of
 $\langle{\cal U}_\sigma D\rangle_c$ :
\begin{eqnarray}\label{DzeroNice}
   D_0={\int{\cal D}^\prime[\sigma]  
   \exp\left(-\chi\beta\sum\limits_{r>0}\mid\sigma_r\mid^2\right)
   \langle{\cal U}_\sigma\rangle_0\langle{\cal U}_\sigma D\rangle_c
   \over
   \int{\cal D}^\prime[\sigma]
   \exp\left(-\chi\beta\sum\limits_{r>0}\mid\sigma_r\mid^2\right)
   \langle{\cal U}_\sigma \rangle_0}
   \;.
\end{eqnarray}
$\langle{\cal U}_\sigma \rangle_0$ is given in (\ref{AverageU}) and 
(\ref{AsubR}), whereas for $\langle{\cal U}_\sigma D\rangle_c$ we expand
\begin{eqnarray}\label{ExpandUD}
   \langle{\cal U}_\sigma D\rangle_c=\langle D\rangle_0
   +{1\over 2}\chi^2\sum\limits_{rs\neq 0}\sigma_r\sigma_s
   \int\limits_0^\beta d\tau d\tau^\prime e^{i\omega_r\tau}
   e^{i\omega_s\tau^\prime} 
   \langle{\rm T}V(\tau)V(\tau^\prime)D(0)\rangle_c
   +{\cal O}(\sigma^3)
   \;.
\end{eqnarray}
 Notice the absence of  first-order terms in $\sigma_r$ since they vanish 
upon the integration in (\ref{DzeroNice}).

Before continuing we introduce the following notation. Using (\ref{DynZ}) and
(\ref{AverageU}) we define a matrix $A(\sigma_0)$ by
\begin{eqnarray}\label{DefineA}
   \zeta_0^\prime=\int{\cal D}^\prime [\sigma]\exp\left[
   -\chi\beta\sum\limits_{r>0}\left(1+a_r\right)\mid\sigma_r\mid^2\right]
   \equiv\int{\cal D}^\prime [\sigma]\exp\left(
   -\chi\beta\sigma^TA\sigma\right)
   \;,
\end{eqnarray}
where the vector $\sigma$ has $N-1$ components
$(\sigma_1^\prime,\sigma_2^\prime,\cdots,\sigma_1^{\prime\prime},
\sigma_2^{\prime\prime},\cdots)$ and $A$ is diagonal with elements
$(1+a_1,1+a_2,\cdots,1+a_1,1+a_2,\cdots)$ with $a_r$ given in (\ref{AsubR}).
We also use (\ref{ExpandUD}) to define $b(\sigma_0)$ and a matrix 
$B(\sigma_0)$ by
\begin{eqnarray}\label{DefineBb}
   \langle{\cal U}D\rangle_c\equiv b+{1\over 2}\sigma^TB\sigma
   \;.
\end{eqnarray}
Using this notation we find
\begin{eqnarray}\label{DzeroABb}
   D_0^{(SPA)}&=&b(\sigma_0)
   \;,\nonumber\\
   D_0^{(PSPA)}&=&b(\sigma_0)
   +{1\over 4\chi\beta}{\rm Tr}\left[A^{-1}(\sigma_0)B(\sigma_0)\right]
   \;.
\end{eqnarray}

Recall that at each $\sigma_0$ we choose a different basis (\ref{StatBasis}) 
for the Fock space in which we write
\begin{eqnarray}\label{StatBasisD}
   D=\sum\limits_{ij}d_{ij}(\sigma_0)
   a_i(\sigma_0)^\dagger a_j(\sigma_0)
   \;.
\end{eqnarray}
The zeroth-order term is then
\begin{eqnarray}\label{bee}
   b(\sigma_0)=\langle D\rangle_0=\sum\limits_i d_{ii}f_i 
   \;.
\end{eqnarray}
In order to obtain an expression for ${\rm Tr}(A^{-1}B)$ in (\ref{DzeroABb}) we 
calculate the double integral in (\ref{ExpandUD}) using Wick's theorem and the 
frequency summation technique \cite{ManyBody,ManyPart}. Notice that 
since A is 
diagonal it is sufficient to consider  the case $s=-r$. The calculation is 
similar to the one performed in the previous Section and yields the result
\begin{eqnarray}\label{TraceAB}
   {\rm Tr}(A^{-1}B)=2\chi^2\sum\limits_{ijk} v_{ij}v_{jk}d_{ki}
   \sum\limits_{r\neq 0}a_r^{-1} I_r^{ijk} 
   \;,
\end{eqnarray}
where 
\begin{eqnarray}\label{Irijk}
   I_r^{ijk}={f_i\over(\Delta_{ij}-i\omega_r)\Delta_{ik}}
   -{f_j\over(\Delta_{ij}-i\omega_r)(\Delta_{jk}+i\omega_r)}
   +{f_k\over\Delta_{ik}(\Delta_{jk}+i\omega_r)}
   \;.
\end{eqnarray}
In the cases $\epsilon_j =\epsilon_i$ etc. it is understood that 
$I_r^{ijk}\equiv\lim\limits_{\epsilon_j\rightarrow\epsilon_i}I_r^{ijk}$.

The sum over $r\neq 0$ in (\ref{TraceAB}) which consists of $N-1$ elements can 
be expressed in a closed form for $N \rightarrow \infty$ by applying once more 
the frequency summation technique. Writing $a_r^{-1}=a^{-1}(i\omega_r )$ and 
$I_r^{ijk}=I^{ijk}(i\omega_r)$, we observe that the sum is carried out over 
the points $z=i\omega_r$ which are the poles of the function 
$\beta/(e^{\beta z}-1)$ with residue of one. Therefore 
\begin{eqnarray}\label{SumToInteg}
   \sum\limits_{r\neq 0}a_r^{-1}(i\omega_r) I_r^{ijk}(i\omega_r)= 
   {1\over{2\pi i}}\oint\limits_{\cal C} dz{\beta\over e^{\beta z}-1} 
    a^{-1}(z) I^{ijk}(z) \;-\; a^{-1}(0) I^{ijk}(0)
   \;,
\end{eqnarray}
where the contour ${\cal C}$ encircles the imaginary axis. 
To evaluate the integral 
we transform ${\cal C}$ into an arbitrarily large circle, deformed 
to include the poles 
at $z=\Delta_{ij}$, $z=-\Delta_{jk}$, and $z=\pm\Omega_\nu$ for all $\nu$,
and  obtain
\begin{eqnarray}\label{DzeroFinal}
   D_0^{(SPA)}&=&\sum\limits_i d_{ii}f_i
   \;,\nonumber\\
   D_0^{(PSPA)}&=&\sum\limits_i d_{ii}f_i
   -{\chi\over 2\beta}\sum\limits_{ijk}v_{ij}v_{jk}d_{ki}
   \left[\sum\limits_\alpha{\beta\over e^{\beta z_\alpha}-1}
   {p_{ijk}(z_\alpha)\over\prod\limits_{\alpha^\prime\neq\alpha} 
                    (z_\alpha-z_{\alpha^\prime})}
   +{p_{ijk}(0)\over\prod\limits_{\alpha^\prime}(-z_{\alpha^\prime})}\right] \;.
\end{eqnarray}
Here
\begin{eqnarray}\label{Pee}
    p_{ijk}(z)&=&\left[{f_k\over\Delta_{ik}}(z-\Delta_{ij})
    -{f_i\over\Delta_{ik}}(z+\Delta_{jk})+f_j\right]
    \prod\limits_{lm}{^\prime}(z^2-\Delta_{lm}^2) \;,\nonumber\\
    z_\alpha&=&\Delta_{ij},-\Delta_{jk},\pm\Omega_\nu
\end{eqnarray}
and the prime in $\prod\limits_{lm}{^\prime}$ restricts the product to pairs
$(l,m)$ satisfying $l<m$ and $\Delta_{lm}\neq 0$. 

 In the case of $q>1$ separable interactions in (\ref{ManyV}) the
calculation is more complicated since the matrix $A(\sigma_0)$ in
(\ref{DzeroABb}) is block-diagonal with blocks of dimension $2q$
(see Eqs. (\ref{AverageUManyV})-(\ref{DynZGaussManyV})). However, 
the general form 
(\ref{TraceAB}) including the summation over $r$ still holds and the
subsequent use of frequency summations to get a closed-form result can 
be carried out in this case as well.

\begin{figure}[p]

\vspace*{1 cm}

\centerline{\epsffile{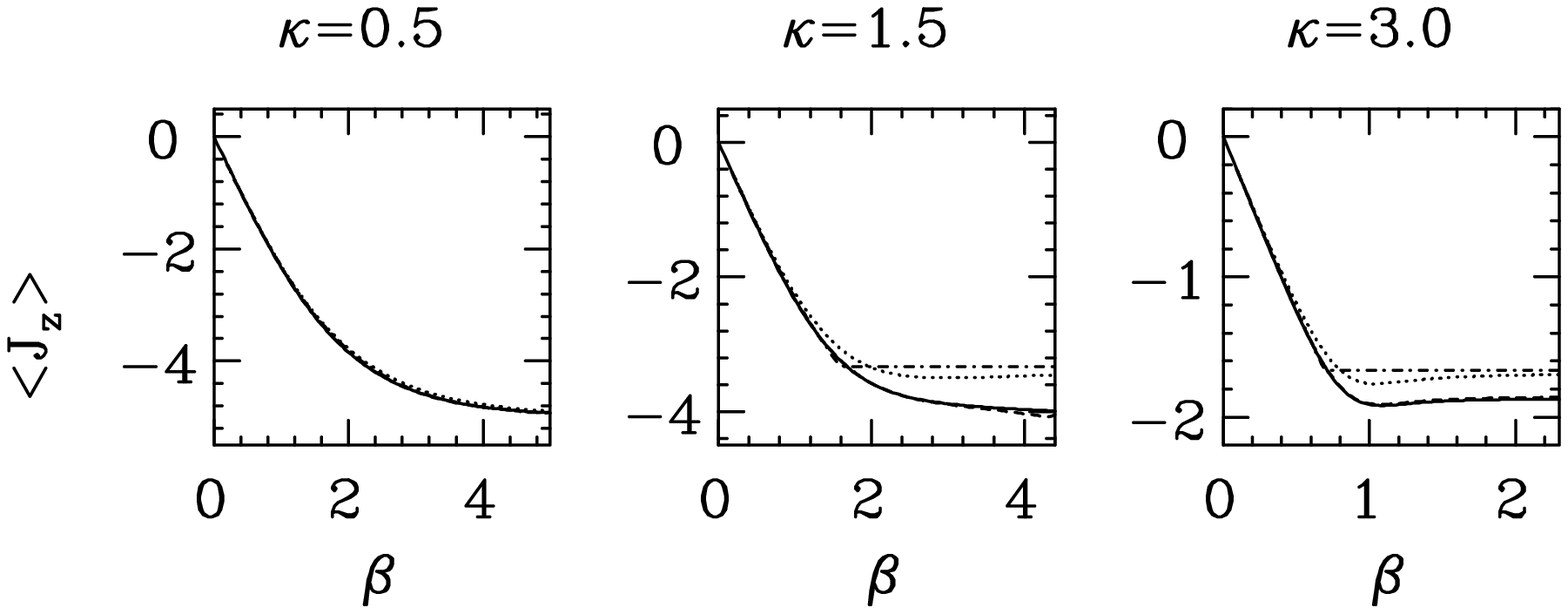}}

\vspace*{-8 cm}

\caption{
Expectation value of $J_z$ as a function of $\beta$ for different values 
of $\kappa$. The SPA result (dotted) is given by
(\protect\ref{Dspa}) and the MFA result (dashed-dotted) is obtained from it
by steepest descent. The PSPA result (dashed) is calculated using 
(\protect\ref{VmatLip})-(\protect\ref{LipkinZ}) and (\protect\ref{DzeroFinal})-
(\protect\ref{DmatLip}). These approximations are compared with the exact result
(\protect\ref{a6}) (solid).
} 
\label{OneB}
\end{figure}

We test the approximations (\ref{DzeroFinal}) by applying them to the
calculation of $\langle J_z\rangle$ in our $U(2)$ model 
($\langle J_x\rangle=\langle J_y\rangle=0$ both exactly and in the above
approximations). The matrix 
corresponding to $D=J_z$ is block-diagonal with $g$ $\; 2\times 2$-blocks
\begin{eqnarray}\label{DmatLip}
   d_{ij}={1\over 2}\left(\begin{array}{c}
   \cos 2\phi \;\;\;\;\;\;\; \sin 2\phi \\
   \sin 2\phi \;\;\;\;\; -\cos 2\phi
   \end{array}\right)
\end{eqnarray}
in the notation of (\ref{VmatLip})-(\ref{Pdiag}). Thus
\begin{eqnarray}\label{Dspa}
   \langle J_z\rangle^{(SPA)}= {1 \over Z^{(SPA)} }\sqrt{\chi\beta\over{2\pi}}
   \int d\sigma_0 e^{-{1\over 2}\chi\beta\sigma_0^2}
   \left(2\cosh{\beta\bar{\epsilon}\over 2}\right)^{2g}
   \times {-g\over 2}\cos 2\phi\tanh{\beta\bar{\epsilon}\over 2}
   \;.
\end{eqnarray}
The more complicated expression for $\langle J_z\rangle^{(PSPA)}$ is similarly
obtained
using (\ref{VmatLip})-(\ref{LipkinZ}) and (\ref{DzeroFinal})-(\ref{DmatLip}).
 The SPA and PSPA
results are presented in Fig.  \ref{OneB}  together with the mean-field
calculation derived from a steepest-descent evaluation of (\ref{Dspa}).
Above the transition temperature all approximations agree fairly well with the 
exact result (\ref{a6}). For lower temperatures the MFA and SPA significantly 
overestimate the exact result whereas the PSPA works well down to 
the breakdown temperature. 

\subsection{Two-Body Observables}
 For simplicity  we assume a two-body observable of the form 
$O=D^\dagger D$ where $D$ is a one-body operator. 
The calculation of the expectation value
of such an observable, albeit more
complicated, can be carried out along the same lines. As for $\langle D\rangle$
we have
\begin{eqnarray}\label{StatDynD2}
   \langle O\rangle={\int d\sigma_0 e^{-\beta F_0} O_0\over
                       \int d\sigma_0 e^{-\beta F_0}} \;,
\end{eqnarray}                    
where
\begin{eqnarray}\label{D2zero}
   O_0={1\over\zeta_0^\prime}\int{\cal D}^\prime[\sigma]  
   \exp\left(-\chi\beta\sum\limits_{r>0}\mid\sigma_r\mid^2\right)
   \langle{\cal U}_\sigma D^\dagger D\rangle_0
   \;.
\end{eqnarray}
Following similar steps we get an expression of
the form (\ref{DzeroABb}) with the same $A(\sigma_0)$ but different 
$B(\sigma_0), b(\sigma_0)$.  After further manipulations we obtain the final 
result
\begin{eqnarray}\label{D2zeroFinal}
   O_0^{(SPA)}&=&b(\sigma_0)
   \;,\nonumber\\
   O_0^{(PSPA)}&=&b(\sigma_0)
   +{\chi\over 4\beta}\sum\limits_{u=1}^6 t_u(\sigma_0) \;,
\end{eqnarray}
where
\begin{eqnarray}\label{BeeTee}
   b(\sigma_0)&=&\sum\limits_{ij}\left[d^\dagger_{ij}d_{ji}f_i(1-f_j)
   +d^\dagger_{ii}d_{jj}f_if_j\right] 
   \;,\nonumber\\
   t_1(\sigma_0)&=&\sum\limits_{ijkl}v_{ij}v_{jk}d_{ki}d^\dagger_{ll}f_l
         S^{(1)}_{ijk} \;,\nonumber\\
   t_2(\sigma_0)&=&\sum\limits_{ijkl}v_{ij}v_{jk}d^\dagger_{ki}d_{ll}f_l
         S^{(1)}_{ijk} \;,\nonumber\\
   t_3(\sigma_0)&=&\sum_{ijkl}v_{ij}v_{kl}d^\dagger_{ji}d_{lk}
         S^{(2)}_{ijkl} \;,\nonumber\\
   t_4(\sigma_0)&=&-\sum\limits_{ijkl}v_{ij}v_{jk}d_{kl}d^\dagger_{li}f_l
         S^{(1)}_{ijk} \;,\nonumber\\
   t_5(\sigma_0)&=&\sum\limits_{ijkl}v_{ij}v_{jk}d^\dagger_{kl}d_{li}(1-f_l) 
         S^{(1)}_{ijk} \;,\nonumber\\
   t_6(\sigma_0)&=&-\sum\limits_{ijkl}v_{ij}v_{kl}d^\dagger_{jk}d_{li}
         S^{(2)}_{ijkl} 
   \;.
\end{eqnarray}
$S^{(1)}_{ijk},S^{(2)}_{ijkl}$ in (\ref{BeeTee}) are given by
\begin{eqnarray}\label{SoneStwo}
   S^{(v)}&=&-\sum\limits_{\alpha}{\beta\over e^{\beta z_\alpha}-1}
   {q^{(v)}(z_\alpha)\over\prod\limits_{\alpha^\prime\neq\alpha}
    (z_\alpha-z_{\alpha^\prime})}
   \;-\;
   {q^{(v)}(0)\over\prod\limits_{\alpha^\prime}(-z_{\alpha^\prime})}
   \;,\;\;\;\;\;
   v=1,2 \;,\nonumber\\
   q^{(1)}_{ijk}(z)&=&p_{ijk}(z) \;\;\; {\rm of} \; (\ref{Pee}) \;,\nonumber\\
   q^{(2)}_{ijkl}(z)&=&-(f_i-f_j)(f_k-f_l)
   \prod\limits_{pq}{^\prime}(z^2-\Delta_{pq}^2) \;,\nonumber\\
   z_\alpha&=&-\Delta_{ij},\Delta_{jk},\pm\Omega_\nu
   \;.
\end{eqnarray}

\begin{figure}[p]

\vspace*{-2 cm}

\centerline{\epsffile{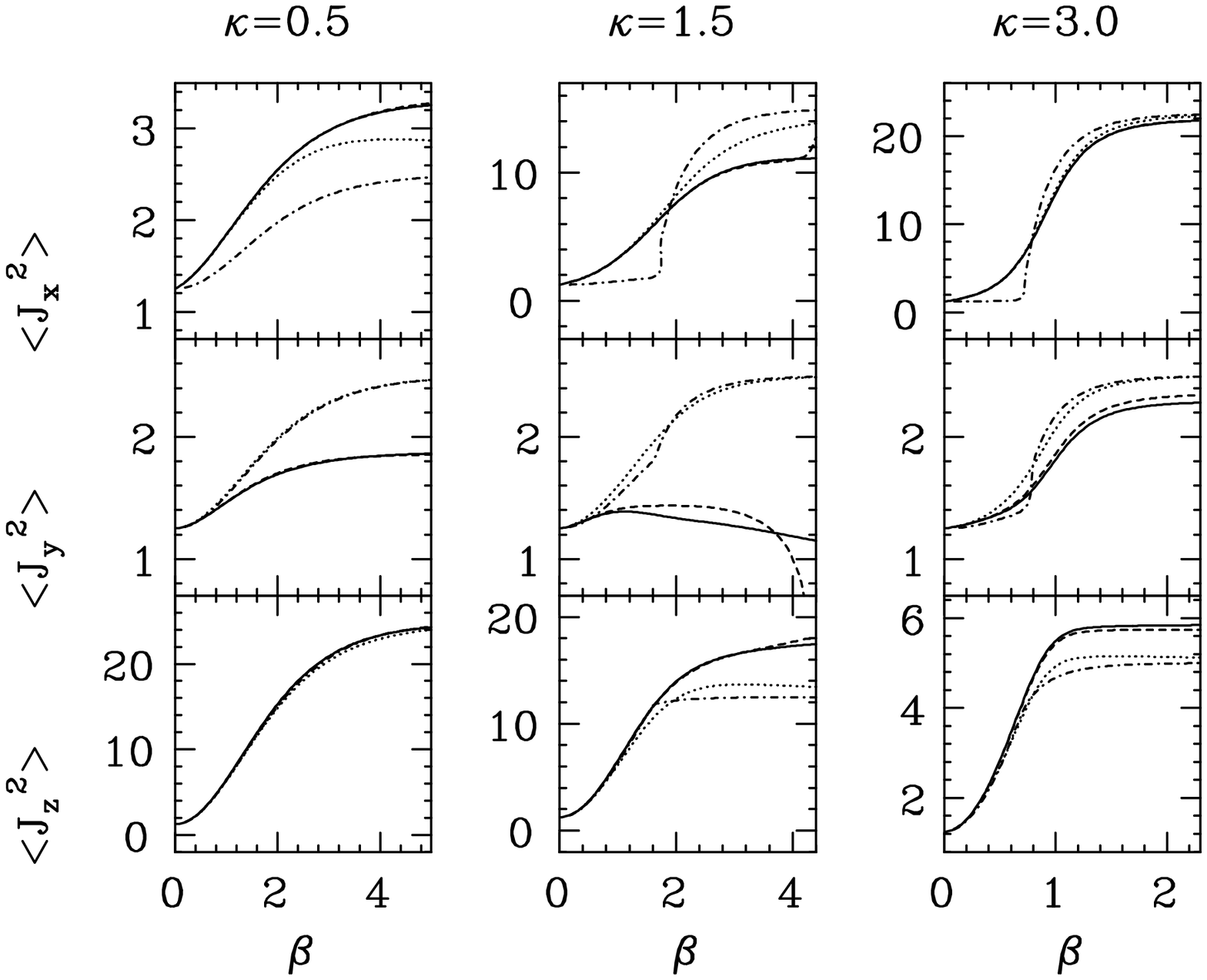}}

\vspace*{-2 cm}

\caption{
Expectation value of $J_x^2$, $J_y^2$ and $J_z^2$ as functions of $\beta$ for 
different values of $\kappa$. The SPA result (dotted) 
is given by (\protect\ref{D2spa}) and the MFA result (dashed-dotted) 
is obtained from it by steepest descent. The PSPA result (dashed) is 
calculated using (\protect\ref{VmatLip})-(\protect\ref{LipkinZ}) and
(\protect\ref{D2zeroFinal})-(\protect\ref{D2matLip}). These approximations are 
compared with the exact result (\protect\ref{a6}) (solid).
} 
\label{TwoB}
\end{figure}

To test the approximations (\ref{D2zeroFinal}) we apply them to the 
calculation of $\langle J_x^2\rangle$, $\langle J_y^2\rangle$ and  
$\langle J_z^2\rangle$ in our $U(2)$ model. The matrices corresponding to 
$D=J_z, J_y$ are block-diagonal with $g$ $\; 2\times 2$-blocks
\begin{eqnarray}\label{D2matLip}
   d_{ij}^{(J_x)}={1\over 2}\left(\begin{array}{c}
   -\sin 2\phi \;\;\;\;\; \cos 2\phi \\
   \;\; \cos 2\phi \;\;\;\;\; \sin 2\phi
   \end{array}\right)
   \;,\;\;\;\;\;
   d_{ij}^{(J_y)}={1\over 2}\left(\begin{array}{c}
   0 \;\;\;\;\; -i \\
   i \;\;\;\;\;\;\; 0
   \end{array}\right)
\end{eqnarray}
whereas $d_{ij}^{(J_z)}$ is given in (\ref{DmatLip}).
We therefore have
\begin{eqnarray}\label{D2spa}
   \langle D^2\rangle^{(SPA)}&=&
  {1 \over Z^{(SPA)}} \sqrt{\chi\beta\over{2\pi}}
   \int d\sigma_0 e^{-{1\over 2}\chi\beta\sigma_0^2}
   \left(2\cosh{\beta\bar{\epsilon}\over 2}\right)^{2g}
   \times (D^2)_0^{(SPA)}
   \;,\nonumber\\
   (J_x^2)_0^{(SPA)}&=&{g^2\over 4}\sin^2 2\phi\tanh^2{\beta\epsilon\over 2}
   +{g\over 8}\left(\cos 4\phi\tanh^2{\beta\bar{\epsilon}\over 2}+1\right)
   \;,\nonumber\\
   (J_y^2)_0^{(SPA)}
   &=&{g\over 8}\left(\tanh^2{\beta\bar{\epsilon}\over 2}+1\right)
   \;,\nonumber\\
   (J_z^2)_0^{(SPA)}&=&{g^2\over 4}\cos^2 2\phi\tanh^2{\beta\epsilon\over 2}
   +{g\over 8}\left(-\cos 4\phi\tanh^2{\beta\bar{\epsilon}\over 2}+1\right)
   \;.
\end{eqnarray}
Expressions for $\langle D^2\rangle^{(PSPA)}$ can similarly be derived
using (\ref{VmatLip})-(\ref{LipkinZ}) and (\ref{D2zeroFinal}-\ref{D2matLip}). 
These results are presented in Fig. \ref{TwoB}  together with mean-field
calculations for different values of $\kappa$. The MFA and SPA exhibit large
deviations from the exact results in most cases even above the transition
temperature and when no transition occurs. The PSPA shows the best agreement 
and 
in the only case where its deviation is appreciable ($\langle J_y^2\rangle$ at 
$\kappa=1.5$), the other approximations give qualitatively wrong results. Note
that the quantitative differences between the various approximations depend on
the observable being calculated. In the no-transition case $\kappa=0.5$ all
approximations agree for $\langle J_z^2\rangle$ (as they do for 
$\langle J_z\rangle$) but not for $\langle J_{x}^2\rangle$, 
$\langle J_{y}^2\rangle$.

\begin{figure}[p]

\vspace*{1 cm}

\centerline{\epsffile{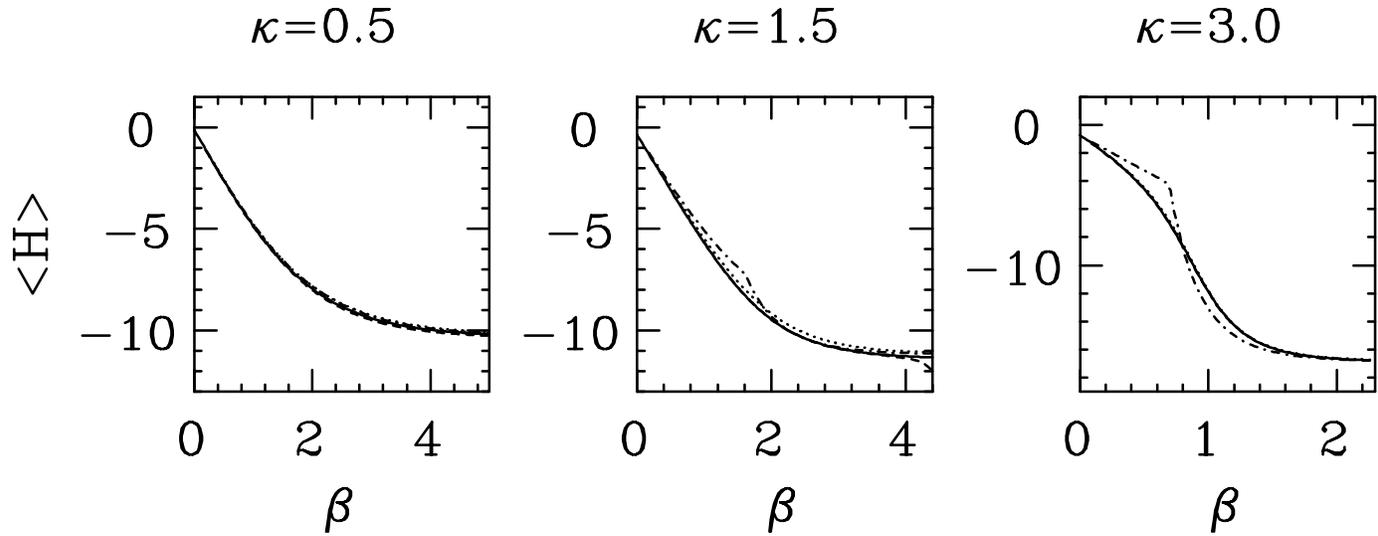}}

\vspace*{-8 cm}

\caption{
Expectation value of the Hamiltonian $H=2\epsilon J_z-2\chi J_x^2$
(\protect\ref{a1}) as a function of $\beta$ for different values of $\kappa$.
Shown are the SPA (dotted), MFA (dashed-dotted), PSPA (dashed) and exact
(solid) results.
} 
\label{Ener}
\end{figure}

The expectation value of the Hamiltonian $H=2\epsilon J_z-2\chi J_x^2$
(\ref{a1}) itself is shown in Fig. \ref{Ener} (note that the additivity of 
the average, $\langle O_1+O_2\rangle=\langle O_1\rangle+\langle O_2\rangle$, 
is preserved in MFA, SPA and PSPA). The PSPA result is still superior but
the agreement of the SPA with the exact result is also quite good, contrary
to what it predicts  for  $\langle J_z\rangle$ and $\langle J_x^2\rangle$ 
when taken separately.

\section{Strength Function}

\subsection{Linear Response Theory}
In this Section we are interested in the response of the system to an external
perturbation
$D$ as reflected by its thermal strength function
\begin{eqnarray}\label{StrengthFn}
   G(\omega)={1\over Z}\sum\limits_{mn}\mid\langle n\mid D\mid m\rangle\mid^2 
   e^{-\beta E_m}\delta(\omega-E_n+E_m) 
   \;,
\end{eqnarray}
where $\mid n\rangle$ and $E_n$ are the many-body eigenstates and 
corresponding 
energies. $G(\omega)$ is the Fourier transform of the real-time response
function $G(t-t^\prime)$ which originates in linear-response theory
\cite{ManyBody,ManyPart} 
\begin{eqnarray}\label{RealTResp}
   G(\omega)&=&\int\limits_{-\infty}^\infty {dt\over 2\pi}
   e^{i\omega t}G(t)
   \;,\nonumber\\
   G(t-t^\prime)&=&-i\langle{\rm T}D_H^\dagger(t)D_H(t^\prime)\rangle
   =-{i\over Z}{\rm Tr}\left[e^{-\beta H}
   {\rm T}D_H^\dagger(t)D_H(t^\prime)\right] 
   \;,
\end{eqnarray}
where the subscript $H$ refers to the Heisenberg picture 
$D_H(t)=e^{iHt}De^{-iHt}$. If an equilibrated system with a Hamiltonian $H$ is 
perturbed by some external potential $V(t)$ at $t=t_0$, the change in the 
thermal expectation value of an observable $D$ with respect to the equilibrium 
situation is
\begin{eqnarray}\label{DeltaDtee}
   \delta\langle D \rangle (t)=-{i\over\hbar}\int\limits_{t_0}^t
   dt^\prime\langle\left[D_H(t),V_H(t^\prime)\right]\rangle\Theta(t-t_0)
   \;.
\end{eqnarray}
Considering an external perturbation of the form 
$V(t)=\alpha(t)D=\sum\limits_{ij}\alpha(t)d_{ij}a_i^\dagger a_j$ 
 we can write (\ref{DeltaDtee}) as
\begin{eqnarray}\label{DeltaDteeRet}
   \delta\langle D(t)\rangle={1\over\hbar}\int\limits_{t_0}^\infty
   dt^\prime G^R(t-t^\prime)\alpha(t^\prime)
   \;,
\end{eqnarray}
where
\begin{eqnarray}\label{RealTRetResp}
   G^R(t-t^\prime)=-i\langle\left[D_H(t),D_H(t^\prime)\right]\rangle
   \Theta(t-t^\prime)
\end{eqnarray}
is the retarded real-time response function. The  Fourier transform of 
the latter is related to that of $\delta\langle D \rangle (t)$ by
\begin{eqnarray}\label{DeltaDomega}
   \delta\langle D\rangle (\omega)={2\pi\over\hbar}\alpha(\omega)G^R(\omega)
   \;,\;\;\;\;\;
   G^R(\omega)={1\over 2\pi}\int\limits_{-\infty}^\infty dt e^{i\omega t}G^R(t)
   \;.
\end{eqnarray}
The strength function can also be obtained directly from $G^R(\omega)$:
\begin{eqnarray}\label{StrengthFnRet}
   G(\omega)=-{1\over\pi}{1\over 1-e^{-\beta\omega}}{\rm Im}G^R(\omega)
   \;.
\end{eqnarray}
This suggests that in order to approximate $G(\omega)$ we
should approximate $G(t)$ or $G^R(t)$ and Fourier-transform the result.
This has been done in \cite{SPAforG} where $G(\omega)$ has been
 obtained from a
static-path calculation of $G(t)$. However, it is very difficult to 
incorporate the contribution of time-dependent paths into the real-time 
formalism.
In contrast, the imaginary-time framework discussed in the previous Sections is 
quite suitable for this task. Thus instead of $G(t)$ we consider the 
imaginary-time response function \cite{ManyBody,ManyPart}
\begin{eqnarray}\label{ImagTResp}
   {\cal G}(\tau-\tau^\prime)=
   -\langle{\rm T}D_H^\dagger(\tau)D_H(\tau^\prime)\rangle
   =-{1\over Z}{\rm Tr}\left[e^{-\beta H}
   {\rm T}D_H^\dagger(\tau)D_H(\tau^\prime)\right] 
\;,
\end{eqnarray}
where $D_H(\tau)= e^{\tau H} D e^{-\tau H}$. 
${\cal G}$ is periodic with period $\beta$ (${\cal G}(\tau+\beta)={\cal G}
(\tau)$), 
hence we are interested in its Fourier coefficients
\begin{eqnarray}\label{ImagGiomega}
   {\cal G}_n=\int\limits_0^\beta d\tau e^{i\omega_n\tau}{\cal G}(\tau) \;,
\end{eqnarray}
where $\omega_n=2\pi n/\beta$ are the Matsubara frequencies \cite{ManyBody}. 
$G^R(\omega)$ and ${\cal G}_n$ are
related by an analytic continuation since their Lehmann representations are
determined by the same weight function $\rho(\omega)$.  Defining
\begin{eqnarray}\label{LehmannRep}
   \Gamma(z)&=&\int d\omega{\rho(\omega)\over z-\omega}   
   \;,\nonumber\\
   \rho(\omega)&=&{1\over Z}\left(1-e^{-\beta\omega}\right)
   \sum\limits_{mn}\mid\langle n\mid D\mid m\rangle\mid^2 
   e^{-\beta E_m}\delta(\omega-E_n+E_m) 
   \;,
\end{eqnarray}
it is easily verified that
\begin{eqnarray}\label{LehmannCont}
   G^R(\omega)=\Gamma(\omega+i\eta)
   \;,\;\;\;\;\;
   {\cal G}_n=\Gamma(i\omega_n)
\end{eqnarray}
as $\eta\rightarrow 0^+$. We can therefore calculate $G^R(\omega)$ by obtaining
an expression for ${\cal G}_n$ with an explicit dependence on $i\omega_n$, then
perform the analytic continuation by formally replacing 
$i\omega_n\rightarrow\omega+i\eta$. We can then calculate
 the strength function $G(\omega)$ from (\ref{StrengthFnRet}). 
 In general, an analytic continuation of
${\cal G}_n$ is not unique since it is based on extrapolating from a discrete
set of points $i\omega_n$ to a continuum $\omega$. The sum rule 
$\int\limits_{-\infty}^\infty d\omega\rho(\omega)=\langle D^\dagger D\rangle$,
however, selects a continuation that satisfies 
$\Gamma(z)\sim\langle D^\dagger D\rangle/z$ as $z\rightarrow\infty$ which
is unique \cite{ManyBody,ManyPart}. In this Section we present an 
auxiliary-field path integral treatment of 
${\cal G}(i\omega_n)\equiv{\cal G}_n$ from which we extract several 
approximations for $G(\omega)$.
                                   
\subsection{Strength Function in the PSPA}   
               
In the auxiliary-field path integral representation Eqs. 
(\ref{ImagTResp})  and (\ref{ImagGiomega}) can be written in the form
\begin{eqnarray}\label{StatDynG}
   {\cal G}(i\omega_n)={\int d\sigma_0 e^{-\beta F_0}{\cal G}_0(i\omega_n)\over
                        \int d\sigma_0 e^{-\beta F_0}} \;,
\end{eqnarray}                    
where
\begin{eqnarray}\label{Gzero}
   {\cal G}_0(i\omega_n)={1\over\zeta_0^\prime}\int{\cal D}^\prime[\sigma]  
   \exp\left(-\chi\beta\sum\limits_{r>0}\mid\sigma_r\mid^2\right)
   \times\int\limits_0^\beta d\tau e^{i\omega_n\tau}
   \langle{\rm T}{\cal U}_\sigma D^\dagger(\tau)D(0)\rangle_0
   \;.
\end{eqnarray}
 The time-dependence in $D$ is understood to be in the interaction picture
with respect to $h_0$, $D(\tau )=e^{\tau h_0}De^{-\tau h_0}$. 
Our aim is to calculate ${\cal G}_0(i\omega_n)$. The connected diagrams factor
out as in (\ref{FactorDconn}) \cite{ManyBody,ManyPart}:
\begin{eqnarray}\label{FactorGconn}
   & &\int\limits_0^\beta d\tau e^{i\omega_n\tau}
   \langle{\rm T}{\cal U}_\sigma D^\dagger(\tau)D(0)\rangle_0
   =\langle{\cal U}_\sigma \rangle_0\int\limits_0^\beta d\tau e^{i\omega_n\tau}
   \left[\langle{\rm T}{\cal U}_\sigma D^\dagger(\tau)D(0)\rangle_c+ 
   \langle{\rm T}{\cal U}_\sigma D^\dagger(\tau)\rangle_c
   \langle{\cal U}_\sigma D(0)\rangle_c\right]
   \nonumber\\
   &=&\langle{\cal U}_\sigma \rangle_0
\int\limits_0^\beta d\tau e^{i\omega_n\tau}
   \langle D^\dagger(\tau)D(0)\rangle_c 
   +{1\over 2}\langle{\cal U}_\sigma \rangle_0
\sum\limits_{rs\neq 0}\sigma_r\sigma_s
   \int\limits_0^\beta d\tau d\tau^\prime d\tau^{\prime\prime}
   e^{i\omega_n\tau}e^{i\omega_r\tau}e^{i\omega_s\tau}
   \nonumber\\ &\times&
   \left[\langle{\rm T}V(\tau^\prime)V(\tau^{\prime\prime})
   D^\dagger(\tau)D(0)\rangle_c 
   +2\langle{\rm T}V(\tau^\prime)D^\dagger(\tau)\rangle_c
   \langle{\rm T}V(\tau^{\prime\prime})D(0)\rangle_c\right]
   \;,
\end{eqnarray}
where terms proportional to $\delta_{n,0}$, such as
\begin{eqnarray}\label{PropToDelta}
   \int\limits_0^\beta d\tau e^{i\omega_n\tau}
   \langle D^\dagger(\tau)\rangle_0\langle D(0)\rangle_0 
   \;,
\end{eqnarray}
are omitted since they do not contribute to the analytic continuation.
Similarly to (\ref{DzeroABb}) we can rewrite ${\cal G}_0(i\omega_n)$
in the form
\begin{eqnarray}\label{GzeroABb}
   {\cal G}_0(i\omega_n)&=&{1\over\zeta_0^\prime}
   \int{\cal D}^\prime[\sigma]  e^{-\chi\beta\sigma^T A\sigma}
   \left[b(i\omega_n)+{1\over 2}\sigma^TB(i\omega_n)\sigma\right]
   \nonumber\\
   &=&b(i\omega_n)+{1\over 4\chi\beta}{\rm Tr}\left[A^{-1}B(i\omega_n)\right]
   \;,
\end{eqnarray}
where $A(\sigma_0)$ is defined in (\ref{DefineA}) and $b(\sigma_0,i\omega_n)$, 
$B(\sigma_0,i\omega_n)$
are given implicitly by (\ref{FactorGconn}). The calculation of these
quantities using Wick's theorem and the frequency summation technique is
similar to those discussed in previous Sections but much more cumbersome
and we give here the final results. We have
\begin{eqnarray}\label{GzeroFinal}
   {\cal G}_0^{(SPA)}(i\omega_n)&=&b(\sigma_0,i\omega_n)
   \;,\nonumber\\
   {\cal G}_0^{(PSPA)}(i\omega_n)&=&b(\sigma_0,i\omega_n)
   +{\chi\over 4\beta}\sum\limits_{u=1}^6 t_u(\sigma_0,i\omega_n)
   \;,
\end{eqnarray}
where
\begin{eqnarray}\label{BeeTeeiomega}
   b(\sigma_0,i\omega_n)&=&-\sum\limits_{ij}d^\dagger_{ij}d_{ji}
   {f_i-f_j\over i\omega_n+\Delta_{ij}}
   \;,\nonumber\\
   t_1(\sigma_0,i\omega_n)&=&
   -2\sum\limits_{ijkl}v_{ij}d^\dagger_{ji}v_{kl}d_{lk}
   {f_i-f_j\over i\omega_n-\Delta_{ij}}
   {f_k-f_l\over i\omega_n+\Delta_{kl}}
   {\prod\limits_{pq}{^\prime}\left[(i\omega_n)^2-\Delta_{pq}^2\right]\over
    \prod\limits_\nu\left[(i\omega_n)^2-\Omega_\nu^2\right]}
   \;,\nonumber\\
   t_2(\sigma_0,i\omega_n)&=&
   -2\sum\limits_{ijkl}v_{jk}v_{kl}d_{li}d^\dagger_{ij}
   S^{(2)}_{ijkl}
   \;,\nonumber\\  
   t_3(\sigma_0,i\omega_n)&=&
   -2\sum\limits_{ijkl} v_{ij}v_{jk}d^\dagger_{kl}d_{li}
   S^{(3)}_{ijkl}
   \;,\nonumber\\
   t_4(\sigma_0,i\omega_n)&=&
   -2\sum\limits_{ijkl} v_{ij}v_{kl}d^\dagger_{ij}d_{li}
   S^{(4)}_{ijkl}
   \;,
\end{eqnarray}  
with
\begin{eqnarray}\label{Integijkl}
   S^{(2)}_{ijkl}&=&\sum\limits_{r\neq 0}a_r^{-1}I_{n-rr}^{ijkl}             
   \;,\nonumber\\
   S^{(3)}_{ijkl}&=&\sum\limits_{r\neq 0}a_r^{-1}I_{-rrn}^{ijkl}             
   \;,\nonumber\\
   S^{(4)}_{ijkl}&=&\sum\limits_{r\neq 0}a_r^{-1}I_{-rnr}^{ijkl}             
   \;,\nonumber\\
   I_{tsr}^{ijkl}
   &=&{f_i\over(\Delta_{ij}+i\omega_t)(\Delta_{ik}+i\omega_{t+s})
               (\Delta_{il}+i\omega_{t+r+s})}
   \nonumber\\
   &-&{f_j\over(\Delta_{ij}+i\omega_t)(\Delta_{jk}+i\omega_s)
               (\Delta_{jl}+i\omega_{r+s})}
     +{f_k\over(\Delta_{ik}+i\omega_{t+s})(\Delta_{jk}+i\omega_s)
               (\Delta_{kl}+i\omega_r)}
   \nonumber\\
   &+&{f_l\over(\Delta_{il}+i\omega_{t+r+s})(\Delta_{jl}+i\omega_{r+s})
               (\Delta_{kl}+i\omega_r)}
   \;.
\end{eqnarray}
The apparent divergences of $I^{tsr}_{ijkl}$, e.g. when both $\Delta_{jl}=0$ and
$r+s=0$, are handled in the usual way by taking the limit 
$\epsilon_j\rightarrow\epsilon_l$. 

We use frequency summations to bring the infinite sums in $S_{ijkl}^{(u)}$ 
into closed forms:
\begin{eqnarray}\label{Stwothreefour}
   S_{ijkl}^{(u)}(\sigma_0,i\omega_n)=-\sum\limits_{\alpha}
   {\beta\over e^{\beta z_\alpha}-1}
   {p^{(u)}_{ijkl}(z_\alpha)\over\prod\limits_{\alpha^\prime\neq\alpha}
    (z_\alpha-z_{\alpha^\prime})}
   \;-\;
   {p^{(u)}_{ijkl}(0)\over\prod\limits_{\alpha^\prime}(-z_{\alpha^\prime})}
   \;,
\end{eqnarray}
where
\begin{eqnarray}\label{Ptwo}
   p^{(2)}_{ijkl}(z)&=&\prod\limits_{pq}{^\prime}(z^2-\Delta_{pq}^2)
   \nonumber\\ &\times&
   \left[-{f_i\over\eta_1\eta_2}(z-z_2)(z-z_3)
         +{f_j\over\eta_1\eta_3}(z-z_1)(z-z_3) 
         +f_k-{f_l\over\eta_2\eta_3}(z-z_1)(z-z_2)\right] 
   \;,\nonumber\\
   z_\alpha&=&\Delta_{ik}+i\omega_n,\Delta_{jk},-\Delta_{kl},\pm\Omega_\nu
   \;, \alpha=1,2,\cdots \;,\nonumber\\
   \eta_1&=&\Delta_{ij}+i\omega_n,\eta_2=\Delta_{il}+i\omega_n,
   \eta_3=\Delta_{jl} 
   \;, 
\end{eqnarray}
\begin{eqnarray}\label{Pthree}
   p^{(3)}_{ijkl}(z)&=&\prod\limits_{pq}{^\prime}(z^2-\Delta_{pq}^2)
   \nonumber\\ &\times&
   \left[-{f_i\over\eta_1\eta_2}(z-z_2)(z-z_3)
         +{f_k\over\eta_1\eta_3}(z-z_1)(z-z_3) 
         +f_j-{f_l\over\eta_2\eta_3}(z-z_1)(z-z_2)\right]
   \;,\nonumber\\ 
   z_\alpha&=&\Delta_{ij},-\Delta_{jk},-\Delta_{jl}-i\omega_n,\pm\Omega_\nu
   \;, \alpha=1,2,\cdots \;,\nonumber\\
   \eta_1&=&\Delta_{ik},\eta_2=\Delta_{il}+i\omega_n,
   \eta_3=\Delta_{kl}+i\omega_n
   \;, 
\end{eqnarray}
\begin{eqnarray}\label{Pfour}
   &p&^{(4)}_{ijkl}(z)=\prod\limits_{pq}{^\prime}(z^2-\Delta_{pq}^2)
   \nonumber\\ &\times&
   \left[{f_i\over\eta_1}(z-z_3)(z-z_4)
        +{f_j\over\eta_2}(z-z_2)(z-z_4) 
        -{f_k\over\eta_2}(z-z_1)(z-z_3)
        -{f_l\over\eta_1}(z-z_1)(z-z_2)\right] 
   \;,\nonumber\\
   z_\alpha&=&\Delta_{ij},\Delta_{ik}+i\omega_n,-\Delta_{jl}-i\omega_n,
   -\Delta_{kl},\pm\Omega_\nu
   \;, \alpha=1,2,\cdots \;,\nonumber\\
   \eta_1&=&\Delta_{il}+i\omega_n,\eta_2=\Delta_{jk}+i\omega_n
   \;.
\end{eqnarray}
Eqs. (\ref{GzeroFinal})-(\ref{BeeTeeiomega}) and 
(\ref{Stwothreefour})-(\ref{Pfour}) 
constitute our final result for ${\cal G}_0(i\omega_n)$. The analytic
continuation (\ref{LehmannCont}) followed by employing the relation 
(\ref{StrengthFnRet}) result in
\begin{eqnarray}\label{GomegaFinal}
   G_0^{(SPA,PSPA)}(\omega)=-{1\over\pi}{1\over 1-e^{-\beta\omega}}
   {\rm Im}{\cal G}_0^{(SPA,PSPA)}(i\omega_n\rightarrow\omega+i\eta)
   \;.
\end{eqnarray}
We remark that the use of frequency summations to convert the infinite sums
in $S_{ijkl}^{(u)}$ (\ref{Integijkl}) into the finite expressions
(\ref{Stwothreefour}) is essential to the extraction of the strength function,
since an analytic continuation of the truncated sums would result in a wrong
functional form of $G(\omega)$.
 
The SPA expression has the simple form
\begin{eqnarray}\label{G0omegaSPA}
   G_0^{(SPA)}(\omega)={1\over 1-e^{-\beta\omega}}
   \sum\limits_{ij}d_{ij}d_{ji}^\dagger(f_i-f_j)\delta(\omega-\Delta_{ij})
   \;.
\end{eqnarray}
This result 
illustrates a significant limitation of this approximation, namely that only
transitions at frequencies $\omega$ corresponding to single-particle energy
differences $\Delta_{ij}(\sigma_0)$ can be described. This shortcoming becomes
evident upon performing the integration over the static field (\ref{StatDynG})
which results in
\begin{eqnarray}\label{GomegaSPA}
   G^{(SPA)}(\omega)={1\over Z^{(SPA)}}{1\over 1-e^{-\beta\omega}}
   \sum\limits_{ij}\sum\limits_\lambda
   \left[e^{-\beta F_0}{d_{ij}d_{ji}^\dagger(f_i-f_j)\over 
   \mid d\Delta_{ij}/d\sigma_0\mid}\right]_{\sigma_0=\sigma_0^\lambda}
\end{eqnarray}
where the second sum is over the values $\sigma_0=\sigma_0^\lambda$
satisfying $\Delta_{ij}(\sigma_0^\lambda)=\omega$. In the case of our
$U(2)$ model $\Delta_{ij}=0, \pm 2\sqrt{\epsilon^2+\chi^2\sigma_0^2}$, hence for
$\mid\omega\mid<2\epsilon$ the strength function $G^{(SPA)}(\omega)=0$ and
the transitions at this $\omega$-range are not reflected.

It is interesting to compare the shell-model Monte Carlo (SMMC)
 methods with the PSPA.
While in  the SMMC  the auxiliary-field path integral
 is evaluated exactly (except for statistical errors),  the problem of 
extracting the strength function 
from the imaginary-time response function is quite difficult 
 due to statistical noise.  The strength function is calculated in the Monte 
Carlo using  a maximal entropy reconstruction method \cite{LG,ShellMC}, 
 but this method work well only in some cases.
  In contrast,  the PSPA  strength function is extracted by
exact analytic continuation (although only within the approximation).
 An additional advantage
of the PSPA is that the infinite-discretization limit of the imaginary-time
interval $[0,\beta)$ is taken exactly, whereas in the SMMC it is necessary 
to extrapolate the finite time step to zero. The validity of the PSPA for 
strength functions is tested in Sections V.C and V.D.

\subsection{Moments of the Strength Function}
The moments of the strength function
\begin{eqnarray}\label{StrengthMom}
   M_n=\int\limits_{-\infty}^\infty d\omega \; \omega^n G(\omega)
   \;,\;\;\;\;\; n=0,1,2,...
\end{eqnarray}
provide another measure of the quality of the approximations we develop.
Rather than integrating over the expressions we have for $G(\omega)$, it is
more convenient to obtain the moments directly in terms of $H$ and $D$. From
$G(\omega)$ being the Fourier transform of $G(t)$ it follows that
\begin{eqnarray}\label{MomAsDeriv}
   M_n=i^n{d^n G\over dt^n}\mid_{t=0}
  \;,
\end{eqnarray}
and differentiating the definition of $G(t)$ (second Eq. in (\ref{RealTResp})) 
we get for the lowest moments
\begin{eqnarray}\label{LowMom}
   M_0&=&\langle D^\dagger D\rangle
   \;,\nonumber\\
   M_1&=&{1 \over 2} \langle \left[ D^\dagger, \left[H,D\right] \right] \rangle
   \;.
\end{eqnarray}
Thus the zeroth moment (total strength) is simply a two-body expectation value,
which was discussed in the previous Section and found to be well reproduced 
 by the PSPA (contrary to the SPA) in our $U(2)$ model 
(see Fig. \ref{TwoB}). 

\begin{figure}[p]

\vspace*{1 cm}

\centerline{\epsffile{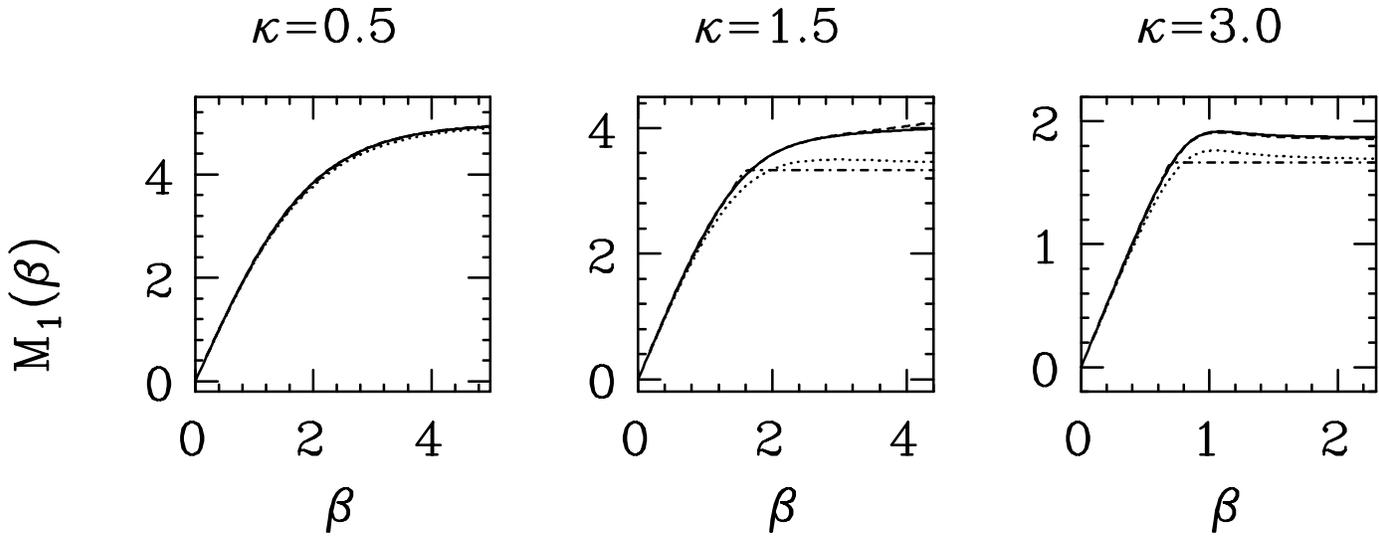}}

\vspace*{-8 cm}

\caption{
First moment of the strength function 
$M_1=\int\limits_{-\infty}^\infty d\omega \; \omega G(\omega)$ as a function of
$\beta$ for different values of $\kappa$, using (\protect\ref{LowMom})
with $D=J_x$. Shown are the SPA (dotted), MFA (dashed-dotted), PSPA (dashed) 
and exact (solid) results.
} 
\label{Smom1}
\end{figure}

The first moment  $M_1$ ($M_1/M_0$ is average transition energy)
 is also given by the expectation value of a two-body
operator. Taking $D=J_x$ in our  $U(2)$ 
model we can exploit the angular-momentum commutation relations to get 
$M_1=2i\epsilon\langle J_x J_y\rangle$. The latter is  
calculated by a  generalization of the results for $\langle D^\dagger D\rangle$ 
obtained in the previous Section to the case
 $\langle D^\dagger_1 D_2\rangle$: whereas 
$d_{ij}$ in (\ref{BeeTeeiomega}) is the matrix corresponding to $D_2$, 
$d_{ij}^\dagger$ should be taken to be that of $D^\dagger_1$. 
The results for $M_1$ 
are shown in Fig. \ref{Smom1} for different values of the mean-field parameter 
$\kappa$ in  (\ref{BetaCLip}). The  PSPA  agrees well with the exact result. 
The SPA is good at high temperatures but worsens appreciably
near the mean-field transition and below it, except for the no-transition case
($\kappa=0.5$) where it remains a good approximation also at low  temperatures.
 This comparison is interesting in
 view of the results of  Ref. \cite{SPAforG} where two versions of the static
approximation for the strength function were tested. The first version 
(called the 
adiabatic approximation) is identical to our SPA. The second version (called the
static-path approximation in Ref. \cite{SPAforG}) consists of estimating 
the path integral 
representation of ${\cal G}(\tau)$ in  (\ref{ImagTResp}) by step function paths
$\xi(\tau^\prime)$ which have a discontinuity at $\tau^\prime=\tau$,
\begin{eqnarray}\label{DiscontSigma}
   \xi(\tau^\prime)=\left\{\begin{array}{c} 
   \sigma_0 \;,\;\;\;\;\; 0\le\tau^\prime<\tau \\
   \rho_0 \;,\;\;\;\;\; \tau\le\tau^\prime<\beta
   \end{array}\right\}
\end{eqnarray}
 rather than by the constant ones 
($\sigma_0=\rho_0$) alone. We shall discuss this approximation further below.
The PSPA includes the effect of discontinuous paths
only to the second order in $\sigma_0-\rho_0$ through the correction factor from
small oscillations about the average value
$\sigma_0\tau/\beta+(\rho_0-\sigma_0)(1-\tau/\beta)$.
It was found in Ref. \cite{SPAforG} that although the inclusion of the 
discontinuous paths (\ref{DiscontSigma}) provided an improvement over the SPA, 
a good agreement with the exact results was not achieved, contrary to the 
situation in the PSPA case. This suggests that the important contribution to 
the moments of the strength function beyond the SPA comes from small-amplitude 
oscillatory paths rather than from step function paths with a large 
discontinuity. 

\subsection{Strength Function in a Simple Model: Results and Discussion}

\begin{figure}[p]

\vspace*{-2 cm}

\centerline{\epsffile{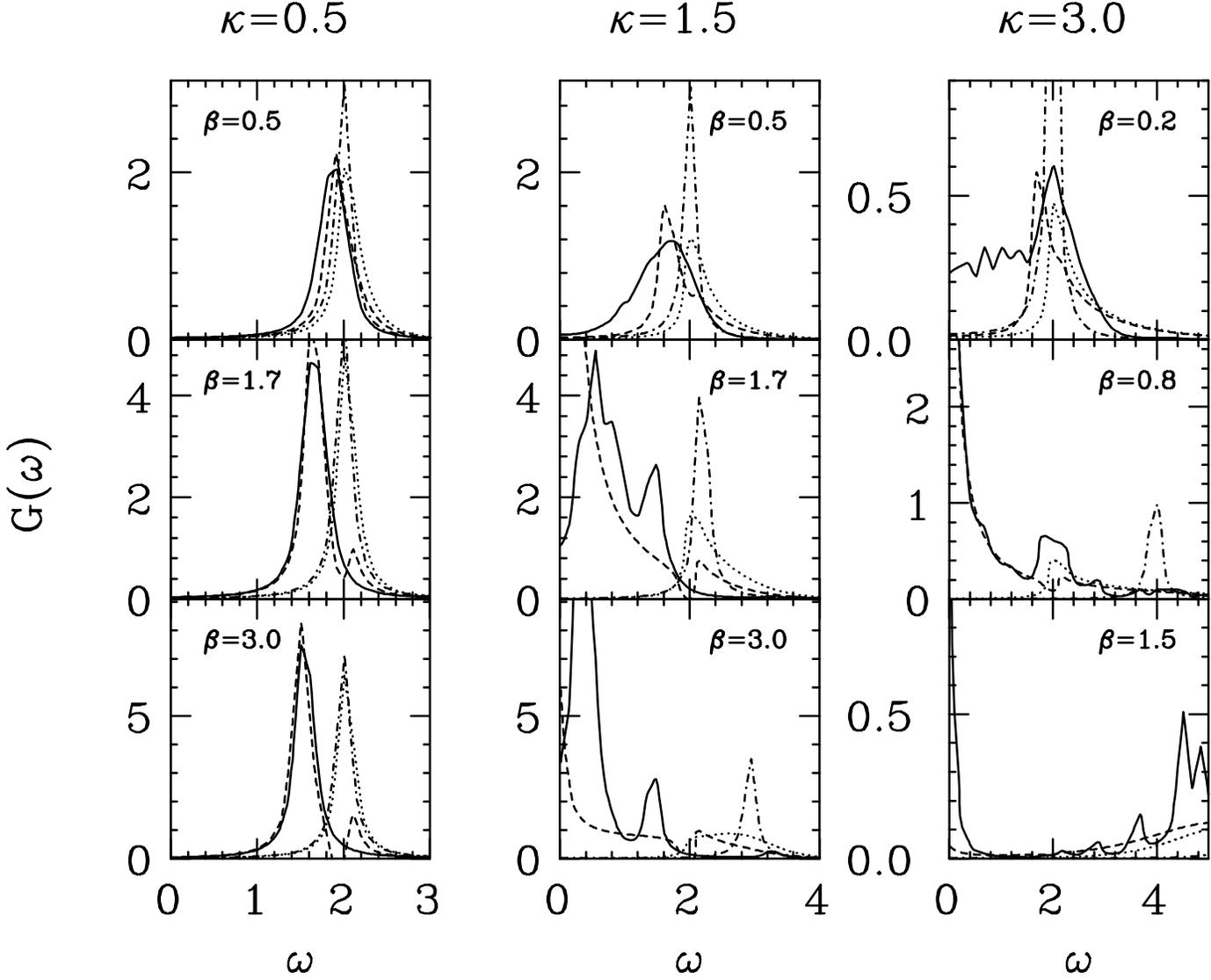}}

\vspace*{-2 cm}

\caption{
Strength function $G(\omega)$ for $D=J_x$ for different values of $\kappa$
and at temperatures above (top), at (middle) and below (bottom) the mean-field
transition (which does not occur for $\kappa=0.5$). Results
are obtained using (\protect\ref{GzeroFinal})-(\protect\ref{BeeTeeiomega}) and 
(\protect\ref{Stwothreefour})-(\protect\ref{GomegaFinal}) with $\eta=0.1$. 
Shown are the SPA (dotted), MFA (dashed-dotted), PSPA (dashed) and exact 
(\protect\ref{a7}) (solid) results. 
} 
\label{Sjx}
\end{figure}

\begin{figure}[p]

\vspace*{-2 cm}

\centerline{\epsffile{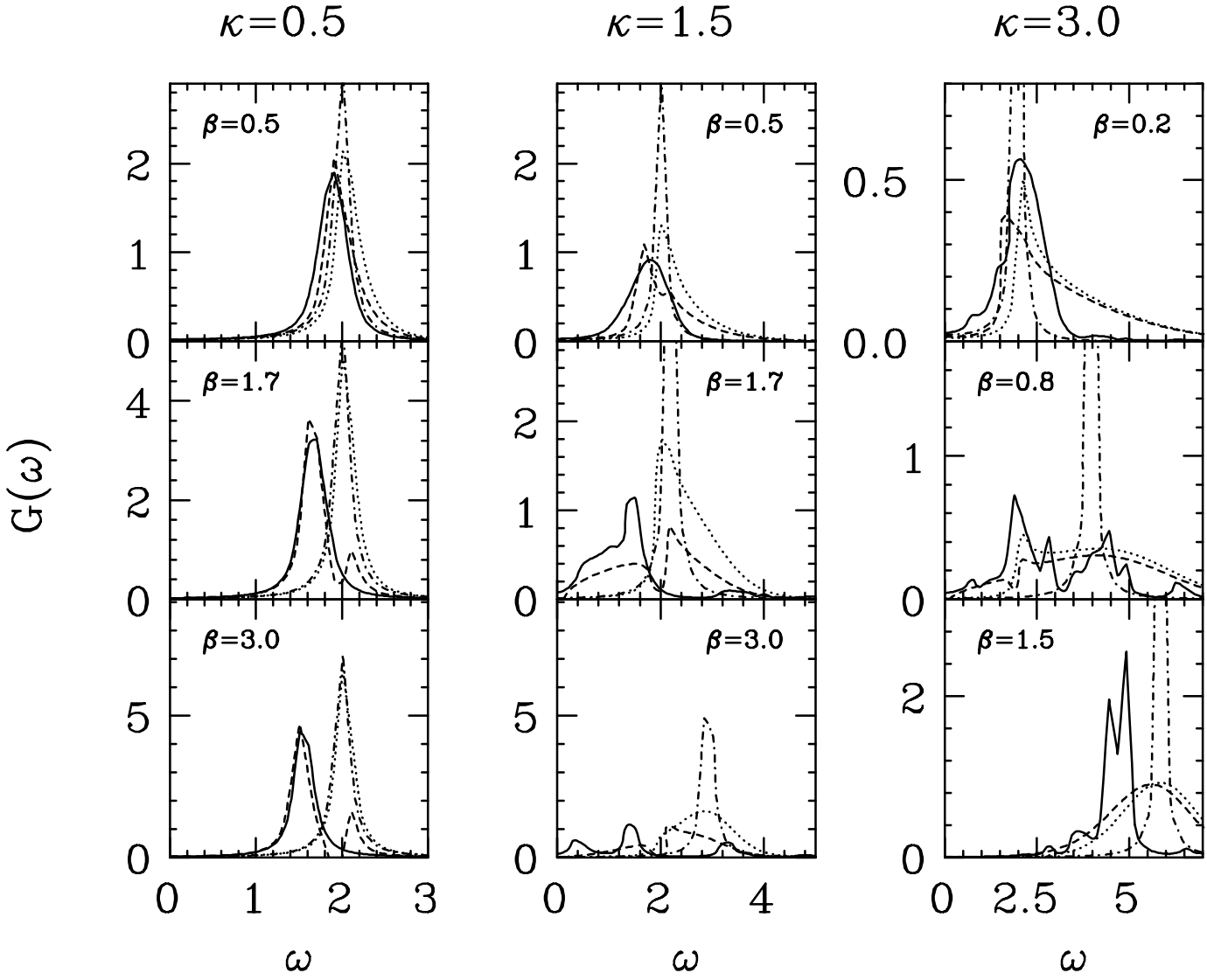}}

\vspace*{-2 cm}

\caption{
Same as in Fig. \protect\ref{Sjx} but for  $D=J_y$. 
} 
\label{Sjy}
\end{figure}

We test the PSPA for $G(\omega)$ in our $U(2)$ model with perturbing
operators $D=J_x$ and $D=J_y$. The results are presented in Fig. \ref{Sjx}
and Fig. \ref{Sjy}, respectively. We consider the same three cases studied 
previously 
characterized by $\kappa=0.5, 1.5, 3.0$ at different temperatures, chosen to
be at the mean-field transition in each case as well as above and below it 
(a transition does not occur for $\kappa=0.5$). Note that the exact result 
(\ref{a7}) consists of a sum of $\delta$-functions in the limit 
$\eta\rightarrow 0^+$ and is therefore singular, 
as is the case for the MFA result. This is not the situation in the SPA and 
PSPA which involve an integration over the static field $\sigma_0$. 
In order to facilitate a meaningful
comparison we keep $\eta$ small but finite both in the exact result, which
becomes a sum of Lorentzians of width $\eta$, and in the approximations.
Finally, we plot $G(\omega)$ only for positive $\omega$ since
$G(-\omega)=e^{-\beta\omega}G(\omega)$ for an Hermitean $D$. 

The shortcoming of the SPA  expressions
 (\ref{G0omegaSPA})-(\ref{GomegaSPA}) is
manifested clearly in Figs. \ref{Sjx} and \ref{Sjy}. Since we use $\epsilon=1$ 
the SPA strength function cannot reflect transitions with $\omega<2$.  Hence
it vanishes in the range  $\omega < 2$ even if most of the strength is
 concentrated there,
as is the case for $\kappa=0.5, 1.5$ at $\beta=1.7, 3.0$ 
(where the SPA result has a shifted peak near $\omega=2$ to the right of the 
exact peak). Furthermore, since the PSPA result consists of an additive 
correction to the SPA expression (see Eq.(\ref{GzeroFinal})),
this shifted SPA peak leaves its trace in the PSPA strength function. For 
$\kappa =0.5$, for instance, even though the main PSPA peak is in excellent
agreement with the exact one, it is accompanies by a 
small additional (false) peak to 
its right, left over from the SPA, which becomes larger as the temperature 
decreases. In general, the PSPA works quite well  for small $\kappa$ but
decreases in quality and becomes comparable with the SPA as $\kappa$
 increases 
or the temperature decreases. The MFA generates a sharp peak located
 near the middle of the broader SPA peak, consistently with  its origin as a 
steepest-descent  approximation of the SPA integral.

We mentioned above that in Ref. \cite{SPAforG}  $G(\omega)$ was calculated 
in a modified
static-path approach which included constant paths with a discontinuity at
$\tau$ in the path integral representation of ${\cal G}(\tau)$. It is
interesting to note that the resulting strength function in \cite{SPAforG}
 is  accurate for large
values of $\kappa$ and deteriorates as $\kappa$ decreases, in contrast with the
PSPA result. This suggests that for small $\kappa$ the major contribution
 to the strength 
function beyond static fields comes from small oscillations about them.
 However, for large $\kappa$  these small oscillations are
negligible and large imaginary-time discontinuities in the static fields become
 important. It is therefore desirable to have an approximation scheme 
which takes both contributions into account. In the following we discuss an
approach to this problem in the imaginary-time framework and the difficulties
it encounters.

We start from the auxiliary-field path integral representation of the 
imaginary-time response function
\begin{eqnarray}\label{ImagTRespPI}
   {\cal G}(\tau)&=&
   -{1\over Z}{\rm Tr}\left[e^{-(\beta-\tau)H}D^\dagger e^{-\tau H}D\right] 
   =-{1\over Z}\int{\cal D}[\xi] \exp\left[-{1\over 2}\chi\int\limits_0^\beta
   d\tau\xi^2(\tau)\right]
   \nonumber\\ 
   &\times&{\rm Tr}\left\{
   {\rm T}\exp\left[-\int\limits_\tau^\beta d\tau^\prime
   \left(K-\xi(\tau^\prime)V\right)\right]D^\dagger
   {\rm T}\exp\left[-\int\limits_0^\tau d\tau^\prime
   \left(K-\xi(\tau^\prime)V\right)\right]D
   \right\}
   \;.
\end{eqnarray}
However,  rather than use the Fourier decomposition (\ref{FourierComp}) of 
$\xi(\tau^\prime)$ over the entire interval $[0,\beta)$ to obtain the form
(\ref{StatDynG}), (\ref{Gzero}) from which the SPA and PSPA are derived, we 
divide the intervals $[0,\tau)$ and $[\tau,\beta)$ into $N$ and $M$ 
sub-intervals, respectively, and use separate decompositions in each. We have
\begin{eqnarray}\label{TwoFourier}
   \xi(\tau^\prime)=\left\{\begin{array}{c} 
   \sum\limits_{r=-(N-1)/2}^{(N-1)/2}\sigma_r e^{i\omega_r\tau^\prime}
   \;,\;\;\;\;\; 0\le\tau^\prime<\tau \\
   \sum\limits_{r=-(M-1)/2}^{(M-1)/2}\rho_r e^{i\nu_r\tau^\prime}
   \;,\;\;\;\;\; \tau\le\tau^\prime<\beta
   \end{array}\right\}
   \;,
\end{eqnarray}
with the reality condition $\sigma_{-r}=\sigma_r^\ast\;,\rho_{-r}=\rho_r^\ast$ 
and $\tau$-dependent frequencies 
$\omega_r=2\pi r/\tau\;,\nu_r=2\pi r/(\beta-\tau)$. In terms of the new
variables the representation (\ref{ImagTRespPI}) becomes
\begin{eqnarray}\label{ImagTResp2Fourier}
   {\cal G}(\tau)
   &=&-{1\over Z}\int{\cal D}[\sigma]{\cal D}[\rho]
   \exp\left[-{1\over 2}\chi\tau\sum\limits_r\mid\sigma_r\mid^2
   -{1\over 2}\chi(\beta-\tau)\sum\limits_r\mid\rho_r\mid^2\right]
   \nonumber\\
   &\times&{\rm Tr}\left\{
   {\rm T}\exp\left[-\int\limits_\tau^\beta d\tau^\prime
   \left(K-\chi\sigma_0 V
   -\chi\sum\limits_{r\neq 0}\sigma_r e^{i\omega_r\tau^\prime}V\right)\right]
   D^\dagger
   \right. \nonumber\\ &\times& \left.
   {\rm T}\exp\left[-\int\limits_0^\tau d\tau^\prime
   \left(K-\chi\rho_0 V
   -\chi\sum\limits_{r\neq 0}\rho_r e^{i\nu_r\tau^\prime}V\right)\right]
   D\right\}
   \;.
\end{eqnarray}
The discontinuous static-path approximation (DSPA), originally introduced in 
\cite{SPAforG} using the real-time framework, is now obtained by neglecting 
the contribution of the oscillations about the static paths
(\ref{DiscontSigma}) which results in the two-dimensional integral 
\begin{eqnarray}\label{GDSPA}
   {\cal G}^{(DSPA)}(\tau)=-{1\over Z^{(DSPA)}} {\chi \over 2\pi} 
\sqrt{\beta(\beta - \tau)}\int d\sigma_0 d\rho_0
   e^{-{1\over 2}\chi\left[\tau\sigma_0^2+(\beta-\tau)\rho_0^2\right]}
   \nonumber\\ 
   \times{\rm Tr}\left[
   e^{-(\beta-\tau)(K-\chi\rho_0 V)}D^\dagger e^{-\tau(K-\chi\sigma_0 V)}D
   \right]
   \;.
\end{eqnarray}
The partition function in this approximation in given by   
\begin{eqnarray}\label{ZDSPA}
   Z^{(DSPA)}=  {\chi \over 2\pi} 
\sqrt{\beta(\beta - \tau)} \int d\sigma_0 d\rho_0
   e^{-{1\over 2}\chi\left[\tau\sigma_0^2+(\beta-\tau)\rho_0^2\right]}
   {\rm Tr}\left[
   e^{-(\beta-\tau)(K-\chi\rho_0 V)}e^{-\tau(K-\chi\sigma_0 V)}
   \right] \;.
\end{eqnarray}
 Note that
$Z^{(DSPA)}$ acquires a $\tau$-dependence.

In order to carry out the imaginary-time technique of Fourier-transforming 
${\cal G}(\tau)$ to get ${\cal G}(i\omega_n)$ and extract the strength
function $G(\omega)$ by an analytic continuation as was done above, it is 
necessary to obtain the functional dependence on $\tau$ in (\ref{GDSPA}) 
analytically. However, the $\tau$-dependence of 
the traces involved is non-trivial and  had to be studied 
numerically in Ref.  \cite{SPAforG} even for 
the simple $U(2)$ model;  note that the analogous situation in the
real-time framework (namely that the $t$-dependence of $G^{(DSPA)}(t)$ is
not given analytically)  does not pose a problem since a
 numerical Fourier transform produces 
$G^{(DSPA)}(\omega)$ directly. Furthermore, unlike the SPA 
case where we had a static Hamiltonian $h_0$  and 
Wick's theorem could be used to calculate the traces,
 here we have a discontinuous  Hamiltonian
\begin{eqnarray}\label{DiscontH}
   h_0(\tau^\prime)=\left\{\begin{array}{c} 
   K-\chi\sigma_0 V  \;,\;\;\;\;\; 0\le\tau^\prime<\tau \\
   K-\chi\rho_0 V \;,\;\;\;\;\; \tau\le\tau^\prime<\beta
   \end{array}\right\}
   \;,
\end{eqnarray}
for which  Wick's theorem is not applicable. In particular, without 
a generalization of Wick's theorem to this situation, it would
 be difficult to use our methods to calculate corrections due to
small oscillations about the discontinuous paths.

\section{Conclusion}

In this paper we present an approximation scheme, the PSPA, for the 
calculation of thermodynamic quantities and finite temperature 
response functions  in finite  fermionic systems. The approximation is 
derived in  the framework of the auxiliary-field path integral. We use an
imaginary-time formulation which facilitates the extension of this approximation
 to physical  quantities beyond the free energy and the 
level density to which it was previously limited. 

Testing the PSPA in a simple many-body model, we find that it improves 
on the SPA and is
a good approximation for  expectation values of observables as well
as for low moments of  strength functions. This  indicates that the 
contribution of
time-dependent fluctuations about the static fields (neglected in the SPA) is 
significant.   The required computational work involved in the PSPA includes
a $q$-dimensional numerical integration over the static fields
$\sigma_0^\alpha$ and a diagonalization of a $q\times q$ matrix at each 
$\sigma_0^\alpha$-point, where $q$ is the number of separable interactions
 in the Hamiltonian. 
This approximation breaks down at low temperatures
 when the small-oscillation
correction factor diverges for the dominant static fields, indicating
 that large time-dependent  fluctuations become
 important. However, the breakdown occurs at temperatures well below
the mean-field transition and does not affect the usefulness of the PSPA 
 except at very low temperatures.
 For the strength function itself the PSPA results become less reliable when 
 the contribution of static paths with large discontinuity (at the time where
 the response function is calculated) is important.
 Further improvement would require the inclusion of
 both discontinuous static paths and small time-dependent oscillations
 around them. 

 It would be interesting to test the PSPA methods for more realistic nuclear 
interactions, such as pairing plus multipole interactions.   
For these interactions, the SPA  works better in a  mixed pairing-density
 decomposition  than in a pure density decomposition. 
Thus,   it would be useful  to extend  the present PSPA techniques 
to such a mixed pairing-density decomposition in the HS representation. 
 
 This work was supported in part by the Department of Energy Grant 
DE-FG02-91ER40608.

\appendix 
\section*{RPA Frequencies at Finite Temperature}

 The finite temperature RPA equations for a general two-body
interaction $u_{ijkl}$ are given by \cite{FTRPA}
\begin{eqnarray}\label{A.1}
-\Delta_{ij} \xi_{ij}^\nu +\sum\limits_{k,l} u_{iljk}(f_k-f_l)\xi_{kl}^\nu
 = \Omega_\nu \xi_{ij}^\nu \;,
\end{eqnarray}
where $\Delta_{ij} = \epsilon_i - \epsilon_j$ and $f_i$ are the Fermi-Dirac
occupation numbers $f_i= (1+ e^{\beta \epsilon_i})^{-1}$.  The solutions of
(\ref{A.1}) are the RPA frequencies  $\Omega_\nu$ and $\xi^\nu$
 are the associated RPA amplitudes. The single-particle
energies $\epsilon_i$ in (\ref{A.1}) correspond to the mean-field solution
 $\bar{\sigma}_0$ but  in the following we replace  $\bar{\sigma}_0$ by 
a general static field $\sigma_0$. 

For a separable interaction as in (\ref{TwoBodyH}) with $V$ Hermitean, 
or more generally 
for an interaction as in (\ref{ManyV}) which is a sum of $q$ such separable 
terms,
 $u_{iljk} = -\sum\limits_\alpha\chi_\alpha v^\alpha_{ij}v^\alpha_{lk}$,
and Eq.  (\ref{A.1}) can be rewritten as
\begin{eqnarray}\label{A.2}
 \xi_{ij}^\nu = -\sum\limits_\alpha\chi_\alpha {v^\alpha_{ij} \over 
\Delta_{ij} + \Omega_\nu} 
\left[ \sum\limits_{kl} v^\alpha_{lk} (f_k-f_l) \xi_{kl}^\nu\right] \;. 
\end{eqnarray}

(\ref{A.2}) can be converted to a set of coupled equations for the $q$ 
quantities 
\begin{eqnarray}\label{A.3}
 \zeta^\nu_\alpha \equiv \sum\limits_{kl} v^\alpha_{lk} 
(f_k-f_l) \xi_{kl}^\nu \; 
\end{eqnarray}
by multiplying  Eq. (\ref{A.3}) by $v^{\alpha^\prime}_{ji} (f_i-f_j)$ and 
summing over $ij$
for each $\alpha^\prime=1, \ldots,q$. We get
\begin{eqnarray}\label{A.4}
\sum\limits_\alpha \left[ \delta_{\alpha^\prime \alpha} + \chi_\alpha  
\sum\limits_{ij}
v_{ji}^{\alpha^\prime}
v_{ij}^\alpha { f_i-f_j \over \Delta_{ij} + \Omega_\nu} \right] 
\zeta^\nu_\alpha =0
\;;\;\;\;\; \alpha^\prime=1,\ldots,q \;.
\end{eqnarray}
For (\ref{A.4}) to have a non-trivial solution (i.e. not all
 $\zeta^\nu_\alpha=0$),
we require that the determinant of the coefficient matrix  vanishes

\begin{eqnarray}\label{A.5}
 \det \left[\delta_{\alpha^\prime \alpha} + \chi_\alpha  \sum\limits_{ij}
v_{ji}^{\alpha^\prime}
v_{ij}^\alpha { f_i-f_j \over \Delta_{ij} + \Omega_\nu} \right] =0\;.
\end{eqnarray}

 Regarding the l.h.s. of Eq. (\ref{A.5}) as a function of $\omega$,  where we 
have substituted $\omega$ for $\Omega_\nu$, we notice
that its roots are  $\pm\Omega_\nu$,
 while its poles are $\pm \Delta_{ij}$. It then follows that

\begin{eqnarray}\label{A.6}
\det \left[\delta_{\alpha^\prime \alpha} + \chi_\alpha  \sum\limits_{ij}
v_{ji}^{\alpha^\prime}
v_{ij}^\alpha { f_i-f_j \over \Delta_{ij} + \omega} \right] 
= {\prod\limits_\nu(\Omega_\nu^2 - \omega^2)
   \over
   \prod\limits_{ij}{^\prime}(\Delta_{ij}^2 - \omega^2)} \;.
\end{eqnarray}
Upon the substitution $\omega \rightarrow i\omega$ we find 
\begin{eqnarray}\label{A.7}
\det \left[\delta_{\alpha^\prime \alpha} + \chi_\alpha  \sum\limits_{ij}
v_{ji}^{\alpha^\prime}
v_{ij}^\alpha { f_i-f_j \over \Delta_{ij} + i\omega} \right] 
= {\prod\limits_\nu(\Omega_\nu^2 + \omega^2)
   \over
   \prod\limits_{ij}{^\prime}(\Delta_{ij}^2 + \omega^2)} \;.
\end{eqnarray}
 Using Eq. (\ref{A.7}) for $\omega=\omega_r$ one obtains  
(\ref{DefOmegaNu})
or  (\ref{DetManyV}) when one or several separable terms in the interaction are
 present, respectively. 

\section*{The Model}
The formalism developed in this paper is illustrated and tested in a simple 
Fermionic system, a variant of a model introduced in \cite{Lipkin} which is 
based on a $U(2)$ algebra and is therefore solvable. This is a  two-level 
system where each level is $g$-fold degenerate and may therefore contain
between zero and $2g$ Fermions. The Hamiltonian is given in terms of 
quasi-angular momentum operators
\begin{eqnarray}\label{a1} 
   H=2\epsilon J_z-2\chi J_x^2
\end{eqnarray}
which has the form (\ref{TwoBodyH}) with $K=2\epsilon J_z$ and $V=2J_x$.
The quasi-angular momentum operators are given by
\begin{eqnarray}\label{JxJz}
   J_x&=&{1\over 2}\sum\limits_{i=1}^g
   \left(a_{1i}^\dagger a_{2i}+a_{2i}^\dagger a_{1i}\right)
   \;,\;\;\;\;\;
   J_y=-{i\over 2}\sum\limits_{i=1}^g
   \left(a_{1i}^\dagger a_{2i}-a_{2i}^\dagger a_{1i}\right)
   \;,\nonumber\\
   J_z&=&{1\over 2}\sum\limits_{i=1}^g
   \left(a_{1i}^\dagger a_{1i}-a_{2i}^\dagger a_{2i}\right)
   \;,\;\;\;\;\;
   \hat{N}=\sum\limits_{i=1}^g
   \left(a_{1i}^\dagger a_{1i}+a_{2i}^\dagger a_{2i}\right)
\end{eqnarray}
with $\hat{N}$ being the particle-number operator. The $2^{2g}$ states are 
arranged in $U(2)$-multiplets $(n,j)$ where the quantum numbers  are the 
number of particles $n=0,...,2g$
and the quasi-angular momentum $j=0,...,n/2$ or $1/2,...,n/2$ 
(depending on whether $n$ is even or odd).
 Each multiplet $(n,j)$ contains $2j+1$ states
labeled by $\mid njm\rangle$ with $m=-j,...,j$ are the eigenvalues of $J_z$. 
To find 
the number $d_n(j)$ of $(n,j)$-multiplets we first observe that the number of 
states with given $n$ and $m$ is
\begin{eqnarray}
  {\cal N}_n(m)=\left(\matrix{g\cr n/2-m}\right)
       \left(\matrix{g\cr n/2+m}\right)
   \;.
\end{eqnarray}
Since every multiplet $(n,j^\prime)$ with $j^\prime\ge j$ contributes a single
state with $J_z=j$, we also have
\begin{eqnarray}
   {\cal N}_n(j)=\sum\limits_{j^\prime=j}^{n/2}d_n(j^\prime)
   \;.
\end{eqnarray}
Therefore
\begin{eqnarray}
   d_n(j)&=&{\cal N}_n(j)-{\cal N}_n(j+1)
   \nonumber\\ &=&
   \left(\matrix{g\cr n/2-j}\right)
   \left(\matrix{g\cr n/2+j}\right)-
   \left(\matrix{g\cr n/2-j-1}\right)
   \left(\matrix{g\cr n/2+j+1}\right)
   \;,
\end{eqnarray}
which checks to give 
\begin{eqnarray}
   \sum\limits_{j=0(1/2)}^{n/2}(2j+1)d_n(j)=\left(\matrix{2g\cr n}\right)
   \;,
\end{eqnarray}
the total number of states with $n$ particles.

The Hamiltonian matrix in this basis is block-diagonal with $d_n(j)$ identical 
blocks of dimension $2j+1$ for each pair $(n,j)$, whose diagonalization gives 
the energies $E_{jm}$ and corresponding eigenstates. The exact partition 
function in the grand canonical ensemble is then
\begin{eqnarray}\label{a5} 
   Z=\sum\limits_{n=0}^{2g}\sum\limits_{j=0(1/2)}^{n/2}d_n(j)
   \sum\limits_{m=-j}^j e^{-\beta(E_{jm}-\mu n)}
   \;.
\end{eqnarray}
The expectation value of an operator $O$ is given by
\begin{eqnarray}\label{a6} 
   \langle O\rangle=
   {1\over Z}\sum\limits_{n=0}^{2g}\sum\limits_{j=0(1/2)}^{n/2}d_n(j)
   \sum\limits_{m=-j}^j\langle jm\mid O\mid jm\rangle e^{-\beta (E_{jm}-\mu n)}
   \;.
\end{eqnarray}
The strength function associated with an operator $O$ is given by
\begin{eqnarray}\label{a7} 
   G(\omega)=
   {1\over Z}\sum\limits_{n=0}^{2g}\sum\limits_{j=0(1/2)}^{n/2}d_n(j)
   \sum\limits_{mm^\prime=-j}^j \mid\langle jm^\prime\mid O\mid jm\rangle\mid^2
   e^{-\beta (E_{jm}-\mu n)}
   \nonumber\\ \times
   {1-e^{-\beta(E_{jm^\prime}-E_{jm})}\over 1-e^{-\beta\omega}}
   {\rm Im}\left[-{1\over\pi}{1\over\omega-(E_{jm^\prime}-E_{jm})+i\eta}\right]
   \;,
\end{eqnarray}
which reduces to a sum over $\delta$-functions as $\eta\rightarrow 0^+$.

\end{document}